\documentclass[prd,twocolumn,nofootinbib,showpacs,superscriptaddress]{revtex4}

\usepackage{amsfonts}
\usepackage{amsmath}
\usepackage{amssymb}
\usepackage{bm}
\usepackage{dcolumn}
\usepackage{graphicx}
\usepackage{graphics}
\usepackage[latin1]{inputenc}
\usepackage{latexsym}
\usepackage{rotating}
\usepackage[colorlinks=true]{hyperref}
\usepackage[all]{hypcap}
\usepackage{xspace} 
\usepackage[usenames]{color}

\usepackage{ulem}
\definecolor {darkgreen}{rgb}{0.2,0.7,0.2}

\newcommand\be{\begin{equation}}
\newcommand\ba{\begin{eqnarray}}
\newcommand\ee{\end{equation}}
\newcommand\ea{\end{eqnarray}}

\newcommand\bw{\begin{widetext}}
\newcommand\ew{\end{widetext}}

\newcommand{\nn}{\nonumber}

\newcommand{\ISCO}{{\mbox{\tiny ISCO}}}

\newcommand{\CS}{{\mbox{\tiny CS}}}

\newcommand{\eff}{{\mbox{\tiny eff}}}
\newcommand{\YP}{{\mbox{\tiny YP}}}
\newcommand{\hor}{{\mbox{\tiny H}}}
\newcommand{\K}{{\mbox{\tiny K}}}
\newcommand{\horK}{{\mbox{\tiny H,K}}}

\newcommand{\Komar}{{\mbox{\tiny Komar}}}

\newcommand{\mrm}{\mathrm}

\begin{document}
\title{Slowly Rotating Black Holes in Dynamical Chern-Simons Gravity: 
\\ Deformation Quadratic in the Spin}

\author{Kent Yagi}
\email{kyagi@physics.montana.edu}
\affiliation{Department of Physics, Montana State University, Bozeman, MT 59717, USA.}

\author{Nicol\'as Yunes}
\affiliation{Department of Physics, Montana State University, Bozeman, MT 59717, USA.}

\author{Takahiro Tanaka}
\affiliation{Yukawa Institute for Theoretical Physics, Kyoto University, Kyoto, 606-8502, Japan.}

\date{\today}

\begin{abstract} 

We derive a stationary and axisymmetric black hole solution to quadratic order in the spin angular momentum. 
The previously found, linear-in-spin terms modify the odd-parity sector of the metric, while the new corrections appear in the even-parity sector. These corrections modify the quadrupole moment, as well as the (coordinate-dependent) location of the event horizon and the ergoregion. Although the linear-in-spin metric is of Petrov type D, the quadratic order terms render it of type I. The metric does not possess a second-order Killing tensor or a Carter-like constant. The new metric does not possess closed timelike curves or spacetime regions that violate causality outside of the event horizon. The new, even-parity modifications to the Kerr metric decay less rapidly at spatial infinity than the leading-order in spin, odd-parity ones, and thus, the former are more important when considering black holes that are rotating moderately fast. We calculate the modifications to the Hamiltonian, binding energy and Kepler's third law. These modifications are crucial for the construction of gravitational wave templates for black hole binaries, which will enter at second post-Newtonian order, just like dissipative modifications found previously. 

\end{abstract}



\maketitle

\section{Introduction}

Although General Relativity (GR) has passed all Solar System and binary pulsar tests~\cite{will-living}, it remains yet to be verified in the non-linear, dynamical strong-field/strong-curvature regime. One of the best astrophysical environments to perform such tests is near black holes (BHs). Mathematical theorems in GR guarantee that the exterior gravitational field of a vacuum, stationary and axisymmetric BH is given by the Kerr metric~\cite{robinson,israel,israel2,hawking-uniqueness0,hawking-uniqueness,carter-uniqueness} 
The multipole moments of an uncharged BH in GR are therefore completely characterized by its mass (mass monopole moment) and spin angular momentum (current dipole moment). If GR is modified, however, BHs need not be represented by the Kerr solution. For example, a stationary BH solution that differs from the Kerr metric has recently been found in quadratic gravity to linear-order in a slow-rotation approximation~\cite{yunespretorius,konnoBH,yunesstein,pani-quadratic}. Future observations of electromagnetic radiation from accretion disks~\cite{psaltis-review,Yunes:2011ws,Kocsis:2011dr} and gravitational radiation from extreme mass-ratio inspirals~\cite{Ryan:1995wh,Ryan:1997hg,Barack:2006pq,glampedakis-emri,emri-review,Sopuerta:2009iy,prisgair} will allow us to probe the spacetime structure and test the Kerr hypothesis: that the massive compact objects at the center of most galaxies are Kerr BHs. 

A well-motivated theory that has recently received much attention is Chern-Simons (CS) modified gravity~\cite{jackiw,CSreview}. The CS action modifies the Einstein-Hilbert one by adding a kinetic scalar field term and a potential composed of the product of this field with the Pontryagin density (the contraction of the Riemann tensor with its dual). Such a potential is necessary to cancel anomalies in heterotic, superstring theory~\cite{polchinski2} and it also arises naturally in loop quantum gravity~\cite{alexandergates,taveras,calcagni} and in effective field theories of inflation~\cite{weinberg-CS}. This theory is also related to 3D topologically massive gauge theories~\cite{deser-TMG}. Two formulations of CS gravity exist: a non-dynamical one and a dynamical one. In the former, the scalar field kinetic term is absent from the action, and thus, this field must be prescribed \textit{a priori}, leading to an extra constraint on the space of solution, i.e. the vanishing of the Pontryagin density~\cite{grumiller}. In the latter, the scalar field kinetic term is kept in the action and the scalar is controlled by a wave equation sourced by the Pontryagin density. Therefore, the dynamical theory preserves diffeomorphism invariance and the strong equivalence principle, although it violates GR's Birkhoff theorem~\cite{grumiller,Yunes:2007ss} and the effacement principle~\cite{quadratic}.     

Dynamical CS gravity is more appealing from a theoretical standpoint but it remains relatively unexplored due to its mathematical complexity. The first study of BH solutions in dynamical CS gravity was carried out by Yunes and Pretorius~\cite{yunespretorius}, and later by Konno \textit{et al.}~\cite{konnoBH}. In these studies, a slow-rotation approximation was used to find the leading-order in spin corrections to the Kerr metric, which enter in the gravitomagnetic sector and modify frame-dragging. A rather weak but robust constraint on dynamical CS gravity was placed~\cite{alihaimoud-chen} with Gravity Probe B~\cite{GPB} (similar constraints can be obtained from LAGEOS and LAGEOS 2~\cite{LAGEOS} through measurements of the Lense-Thirring effect~\cite{iorio-LT}.) Dynamical CS gravity can also be constrained by table-top experiments that confirm Newton's law to length scales above $\mathcal{O}(10\mu$m)~\cite{kapner}. Interestingly, table-top experiments place similar constraint to those obtained from Solar System observations. 

A much stronger constraint can be placed with future gravitational wave observations of compact binary coalescences. This necessitates a calculation of both \textit{dissipative} corrections to the waveform, i.e.~modifications to the rate of change of the energy, angular momentum and Carter constant, and \textit{conservative} corrections, i.e.~modifications to the binding energy and Kepler's Third Law. In~\cite{quadratic}, we calculated the former and found that they introduce modifications to the waveform at 2nd post-Newtonian (PN) order in terms of its frequency (or velocity) dependence, although its amplitude is of course suppressed relative to GR by the CS coupling constant. In order to obtain the full 2PN waveform, we must first compute the 2PN conservative corrections, which require a calculation of (i) a stationary, axisymmetric BH solution at quadratic order in spin, and (ii) the dipole-dipole scalar force acting on the binary components. In this paper, we concentrate on the former. 

We will search for a slowly-rotating BH solution in dynamical CS gravity at quadratic order in spin with two approximations: slow-rotation and small coupling~\cite{yunespretorius}. The former assumes that the BH spin angular momentum is much smaller than its mass squared. The latter postulates that the deformation away from GR is small, which corresponds to a dynamical CS dimensionless coupling constant [defined later in Eq.~\eqref{action}] much smaller than unity. This is a reasonable approximation, given that GR has already passed stringent tests, albeit in the weak field. Moreover, the dynamical CS action is an {\emph{effective theory}} that derives from a leading-order truncation in the dynamical CS coupling parameter of a more fundamental one. Therefore, the action and its associated modified field equations are only valid to leading-order in the dynamical CS coupling parameter. If one did not use the small-coupling approximation to iteratively solve the modified field equations, third time derivatives could arise in the field equations, probably signaling the presence of ghost modes. Presumably, these ghost modes might be eliminated by an appropriate UV completion.

The method we employ to find a slowly-rotating solution at second order in the spin is rather novel in this context. We follow closely BH perturbation theory techniques~\cite{zerilli,sago}: we treat the second-order in spin correction to the dynamical CS metric as a perturbation away from the leading-order in spin solution found by Yunes and Pretorius~\cite{yunespretorius}. We then derive master equations governing this perturbation and decompose the solution in tensor spherical harmonics, which decouples the system into a linear ordinary differential one. The solution to this system is then mathematically straightforward, although we do verify that the full metric satisfies the dynamical CS field equations explicitly with symbolic manipulation software. As expected, the quadratic-order in spin corrections modify only the even-parity sector of the metric.

The properties of this solution are also quite interesting. Since we obtain corrections to quadratic order in the spin, we are able to compute the gauge-dependent shift in the location of the event horizon and the ergosphere, as well as the deformation in the Kerr quadrupole moment. We also find that the new metric retains its Lorentz signature and closed timelike curves do not exist outside of the event horizon. Therefore, such a spacetime is perfectly suitable to the study of photon trajectories when considering the shadow of BH accretion disks. We also consider test-particles motion in this new background. We obtain the corrections to the relation among the binding energy, the (z-component of) angular momentum, and the orbital frequency, and the frequency at the innermost stable circular orbit. 

Perhaps one of the most interesting, albeit not surprising properties of the new solution is that, although the linear order in spin metric~\cite{yunespretorius} remains of Petrov type D~\cite{sopuertayunes}, the quadratic order terms calculated here render the metric of type I. Obviously, this is drastically different from the Kerr metric, which remains of Petrov type D to all orders in a small spin expansion. This is because if a metric is not of Petrov type Dat some order in spin, this violation cannot be compensated for by higher order spin contributions, as long as the small spin approximation holds. The new BH solution does not possess a non-trivial second-rank Killing tensor, and thus, there are no naive extensions of the Carter-like constant and a separable structure is not admissible~\cite{benenti,yasui}. The latter statement means that there does not exist any \textit{spacetime} coordinate transformation that leads to separable Hamilton-Jacobi equations. One cannot, however, rule out a \textit{canonical} transformation that does render the equations separable. This is related to the fact that there might exist a higher-rank Killing tensor that kneads to a fourth constant of the motion. Given this, it is not clear whether geodesics in this new background will be chaotic or not. Nonetheless, upon orbit-averaging, a modified Carter-like constant reappears, which might indicate that orbits are regular, except for resonant ones.

The remainder of this paper presents further details and it is organized as follows. 
Section~\ref{sec:CS} presents the basic equations of dynamical CS gravity. 
Section~\ref{sec:BHsol} begins by describing the approximation used to find BH solutions in this theory and then continues to describe the solution found in~\cite{yunespretorius} and~\cite{konnoBH} and the new solution found in this paper. 
Section~\ref{sec:properties} investigates the basic properties of this new BH solution, such as the (gauge-dependent) location of the event horizon and ergosphere. 
Section~\ref{sec:geodesic} discusses geodesics in this new background and the Petrov type of the new solution. 
Section~\ref{sec:Discussions} summarizes and describe several possible avenues for future research. 

All throughout, we use the following conventions, following mostly Misner, Thorne and Wheeler~\cite{MTW}. We use the Greek letters $(\alpha, \beta, \cdots)$ to denote spacetime indices. The metric is denoted $g_{\mu \nu}$ and it has signature $(-,+,+,+)$. We use geometric units, with $G=c=1$.

\section{Dynamical Chern-Simons Gravity}
\label{sec:CS}

In this section, we introduce the basic equations of dynamical CS gravity and establish notation. The action is defined by~\cite{CSreview}
\ba
S &\equiv & \int d^4x \sqrt{-g} \Big\{ \kappa_g R + \frac{\alpha}{4} \vartheta R_{\nu\mu \rho \sigma} {}^* R^{\mu\nu\rho\sigma}  \nn \\
& &  - \frac{\beta}{2} [\nabla_\mu \vartheta \nabla^{\mu} \vartheta + 2 V(\vartheta) ] + \mathcal{L}_\mrm{mat} \Big\}\,.
\label{action}
\ea
Here, $\kappa_g \equiv (16\pi)^{-1}$, $g$ denotes the determinant of the metric $g_{\mu\nu}$ and $R_{\mu\nu \delta \sigma}$ is the Riemann tensor. 
${}^* R^{\mu\nu\rho\sigma}$ is the dual of the Riemann tensor, defined by~\cite{CSreview}
\be
{}^* R^{\mu\nu\rho\sigma} \equiv \frac{1}{2} \epsilon^{\rho\sigma\alpha\beta} R^{\mu\nu}{}_{\alpha\beta}\,,
\ee
where $\epsilon^{\mu\nu\alpha\beta}$ is the Levi-Civita tensor.
$\vartheta$ is a scalar field, while $\alpha$ and $\beta$ are coupling constants.
$V(\vartheta)$ is an additional potential and $\mathcal{L}_\mrm{mat}$ denotes the matter Lagrangian density.
Following~\cite{yunespretorius}, we take $\vartheta$ and $\beta$ to be dimensionless and $\alpha$ to have dimensions of (length)$^2$.
For convenience, we define a dimensionless parameter 
\be
\zeta \equiv \frac{\alpha^2}{\kappa_g \beta M^4}\,,
\label{zeta}
\ee
where $M$ is the typical mass of the system (or strictly speaking, the curvature length scale).

The field equations in this theory are given by~\cite{CSreview}
\be
G_{\mu\nu} + \frac{\alpha}{\kappa_g} C_{\mu\nu} =\frac{1}{2\kappa_g} (T_{\mu\nu}^\mrm{mat} + T_{\mu\nu}^\vartheta)\,,
\label{field-eq}
\ee
where $G_{\mu\nu}$ is the Einstein tensor and $T_{\mu\nu}^\mrm{mat}$ is the matter stress-energy tensor.
The C-tensor and the stress-energy tensor for the scalar field are defined by
\begin{align}
C^{\mu\nu} & \equiv  (\nabla_\sigma \vartheta) \epsilon^{\sigma\delta\alpha(\mu} \nabla_\alpha R^{\nu)}{}_\delta + (\nabla_\sigma \nabla_\delta \vartheta) {}^* R^{\delta (\mu\nu) \sigma}\,, \\
T_{\mu\nu}^\vartheta & \equiv  \beta (\nabla_\mu \vartheta) (\nabla_\nu \vartheta) -\frac{\beta}{2} g_{\mu\nu} \left[ \nabla_\delta \vartheta \nabla^\delta \vartheta +2V(\vartheta)\right]\,.
\end{align}
Equivalently, Eq.~\eqref{field-eq} can be rewritten as
\be
R_{\mu\nu} =- \frac{\alpha}{\kappa_g} C_{\mu\nu} + \frac{1}{2\kappa_g} \left( \bar{T}_{\mu\nu}^\mrm{mat} +  \bar{T}_{\mu\nu}^{\vartheta} \right)\,,
\label{mod-ein}
\ee
where we introduced the trace-reversed stress-energy tensors 
\begin{align}
\bar{T}_{\mu\nu}^\mrm{mat} & \equiv  T_{\mu\nu}^\mrm{mat} - \frac{1}{2} g_{\mu\nu} T^\mrm{mat}\,, 
\\
\bar{T}_{\mu\nu}^{\vartheta} & \equiv  \beta (\nabla_\mu \vartheta) (\nabla_\nu \vartheta)\,,
\end{align}
and used the fact that the C-tensor is traceless. The evolution equation of the scalar field is given by
\be
\square \vartheta = -\frac{\alpha}{4 \beta} R_{\nu\mu \rho \sigma} {}^* R^{\mu\nu\rho\sigma} + \frac{dV}{d\vartheta}\,.
\label{scalar-wave-eq}
\ee

In 4D, low-energy effective string theories, the (axion) scalar field has a shift symmetry, i.e.~the equations of motion are invariant under the symmetry transformation $\vartheta \to \vartheta + c$, with $c$ a constant, which disallows any mass terms in the action. If one forgets about shift-symmetry and insists on a mass term, then the dynamics of the scalar field would freeze and their would not be much scalar-field propagation. In dynamical CS gravity, however, such mass terms are not allowed. For simplicity, in this paper we set $V(\vartheta)=0$ throughout. 

Dynamical CS gravity should be thought of as an effective theory, and as such, it possess a cut-off scale outside which its action should be modified through the inclusion of higher-order curvature terms. This cut-off scale can be determined by estimating the order of magnitude of loop corrections to the second term in Eq.~\eqref{action} due to $n$-point interactions. Denote the additional number of vertices and scalar and graviton propagators relative to tree level diagrams as $V$, $P_s$ and $P_g$, one has
\be
P_s = \frac{V}{2}, \quad P_g = (n-1)\frac{V}{2}\,.
\ee
One immediately sees that loop corrections are suppressed by a factor of $\alpha^{V} M_\mrm{pl}^{(2-n)V} \Lambda^{nV}$, where $M_\mrm{pl}$ is the Planck mass and $\Lambda$ is the energy scale introduced such that the suppression factor becomes dimensionless. This factor becomes of order unity when $\Lambda$ takes the critical value
\be
\Lambda_c \equiv M_\mrm{pl}^{1-2/n} \alpha^{1/n}\,,
\label{lambdac}
\ee
which corresponds to the cutoff energy scale above which one cannot treat dynamical CS gravity as an effective theory. 

Given this cut-off scale, one can estimate the value of $\alpha$ above which the theory is not an effective one anymore. First, notice that for a fixed value of $\alpha$, $\Lambda_c$ becomes larger as $n$ increases. Hence, $n=3$ will lead to the most stringent constraint on $\alpha$. From Eq.~\eqref{lambdac}, the critical wavelength scale below which the strong coupling effect cannot be neglected is given by
\be
\lambda_c \equiv L_\mrm{pl}^{1/3} \alpha^{1/3}\,,
\ee
where $L_\mrm{pl}$ is the Planck length scale. When one takes $\lambda_c$ to be of $\mathcal{O}(10\mu$m), thus saturating the table-top experiment bound~\cite{kapner}, one finds that 
\be
\sqrt{\alpha} < \mathcal{O}(10^8 \mrm{km}).
\label{alpha-ineq}
\ee
For values of $\alpha$ that satisfy the above inequality, dynamical CS gravity can be treated as an effective theory and higher-order curvature terms in the action can be neglected. Notice that this inequality is of the same order as the constraint obtained from Solar System experiments~\cite{alihaimoud-chen}. Of course, one can have a value of $\alpha$ that satisfies this inequality, without necessarily having $\zeta \ll 1$. In this paper, however, we are interested in BHs of masses in the range $(3,10^{7}) M_{\odot}$, for which the small coupling approximation requires $\sqrt{\alpha} \ll 10^{7} \mrm{km}$, thus automatically satisfying the inequality in Eq.~\eqref{alpha-ineq}.

\section{Rotating Black Hole Solutions}
\label{sec:BHsol}

In this section, we describe the two approximation schemes that we use to obtain the slowly rotating BH solution in dynamical CS gravity at quadratic order in spin. We then describe the slowly-rotating BH solution at linear-order in spin, found by~\cite{yunespretorius} and~\cite{konnoBH}, and apply these approximations to find the second-order in spin solution.

\subsection{Approximation Schemes}

Following~\cite{yunespretorius}, we consider stationary and axisymmetric BH solutions in dynamical CS gravity with small-coupling ($\zeta \ll 1$) and slow-rotation ($\chi \ll 1$). The small-coupling approximation implies that we consider small CS deformations away from GR.  The metric can then be expanded as
\be
g_{\mu\nu} = g_{\mu\nu}^{(0)} + \alpha'^2 g_{\mu\nu}^{(2)} + \mathcal{O}(\alpha'^4)\,,
\ee
where $\alpha'$ is a bookkeeping parameter that labels the order of the small-coupling approximation, with $g_{\mu \nu}^{(n)} \propto \alpha^n$.
Notice that only terms with even powers in $\alpha'$ appear in the metric. 
Then, we expand each $g_{\mu\nu}^{(0)}$ and $g_{\mu\nu}^{(2)}$ in a slow-rotation expansion via
\ba
g_{\mu\nu}^{(0)} &=& g_{\mu\nu}^{(0,0)} + \chi' g_{\mu\nu}^{(1,0)} + \chi'{}^2 g_{\mu\nu}^{(2,0)} + \mathcal{O}(\chi'^3)\,, \\
\alpha'^2 g_{\mu\nu}^{(2)} &=& \alpha'^2 g_{\mu\nu}^{(0,2)} + \alpha'^2  \chi' g_{\mu\nu}^{(1,2)} + \alpha'^2 \chi'{}^2 g_{\mu\nu}^{(2,2)} \nn \\
& & + \mathcal{O}(\alpha'^2 \chi'^3)\,, 
\label{metric-expand}
\ea
where $\chi'$ is another bookkeeping parameter that labels the order of the slow-rotation approximation.
Notice that $g_{\mu\nu}^{(m,n)} \propto \chi^m \alpha^n$, where $\chi \equiv a/M$ is the dimensionless spin parameter. 

The quantities $g_{\mu\nu}^{(0,0)}$, $g_{\mu\nu}^{(1,0)}$ and $g_{\mu\nu}^{(2,0)}$ can be obtained by expanding the Kerr solution in $\chi \ll 1$, whose line element in Boyer-Lindquist (BL) coordinates $(t,r,\theta,\phi)$ is
\ba
ds_{\K}^2  &= & -\left( 1-\frac{2Mr}{\Sigma} \right) dt^2 - \frac{4Mar\sin^2\theta}{\Sigma}dtd\phi + \frac{\Sigma}{\Delta} dr^2 \nn \\
& & + \Sigma d\theta^2 + \left( r^2+a^2 + \frac{2Ma^2 r \sin^2\theta}{\Sigma}  \right) \sin^2\theta d\phi^2\,, \nn \\
\label{Kerr-metric}
\ea
where $\Delta$ and $\Sigma$ are defined by
\ba
\Delta & \equiv & r^2 - 2Mr + a^2\,, \\
\Sigma & \equiv & r^2 + a^2 \cos^2\theta\,.
\ea
Here, $M$ is the mass of the BH and $a \equiv S/M$ with $S$ denoting the magnitude of spin angular momentum of the BH.

Let us now expand the scalar field $\vartheta$. From Eq.~\eqref{scalar-wave-eq}, we see that the leading-order contribution to $\vartheta$ is proportional to $\alpha$. Therefore, we can expand $\vartheta$ as
\be
\vartheta = \alpha' \left[ \chi' \vartheta^{(1,1)}+ \chi'^2\vartheta^{(2,1)} \right] +\mathcal{O}(\alpha' \chi'^3)\,.
\ee
There is no $\vartheta^{(0,1)}$ term here because the Pontryagin density vanishes when evaluated on spherically symmetric spacetimes. There is no $\mathcal{O}(\alpha'^2)$ term and we have also here neglected terms of ${\cal{O}}(\alpha'^{3})$ since they do not affect the metric perturbation at $\mathcal{O}(\alpha'^2)$. 

\subsection{BH solutions to $\mathcal{O}(\alpha'^{2} \chi')$}

Let us first concentrate on solutions at ${\cal{O}}(\alpha'^{2} \chi'^{0})$. As already mentioned, the Pontryagin density vanishes for any spherically symmetric spacetime~\cite{grumiller}. Thus, static, spherically symmetric BHs are still described by the Schwarzschild solution. This implies that $g_{\mu\nu}^{(0,n)}=0$ for all $n$, and in particular, $g_{\mu \nu}^{(0,2)} = 0$. 

To ${\cal{O}}(\alpha'^{2} \chi')$ in metric, Yunes and Pretorius found that in BL-type coordinates (the coordinates where the GR part of the BH metric is identical to Kerr in BL coordinates)~\cite{yunespretorius,konnoBH} 
\be
\vartheta^{(1,1)} = \frac{5}{8} \frac{\alpha}{\beta} \chi \frac{\cos\theta}{r^2} \left( 1+2\frac{M}{r} + \frac{18}{5} \frac{M^2}{r^2} \right)
\label{theta-linear}
\ee
and the only non-vanishing term in $g_{\mu \nu}^{(1,2)}$ is 
\be
g_{t\phi}^{(1,2)} = \frac{5}{8}\zeta M \chi \frac{M^4}{r^4} \left( 1+\frac{12}{7}\frac{M}{r} + \frac{27}{10} \frac{M^2}{r^2} \right) \sin^2 \theta\,,
\label{YP-sol}
\ee
with all other components set to zero. Therefore, the line-element to ${\cal{O}}(\alpha'^2 \chi')$ is given by
\begin{align}
ds^{2} &= ds^{2}_{K} + \frac{5}{4}\zeta M \chi \frac{M^4}{r^4} \left( 1+\frac{12}{7}\frac{M}{r} + \frac{27}{10} \frac{M^2}{r^2} \right) \sin^2 \theta dt d\phi\,.
\label{metric-linear}
\end{align}
Notice that although the correction term does not diverge at the unperturbed Schwarzschild horizon, $r=2M$, it does not vanish there, either. One can, however, resum the metric such that the correction indeed vanishes at the Schwarzschild horizon, as we will discuss in Sec.~\ref{sec:solution}.

\subsection{BH solutions at $\mathcal{O}(\alpha'^{2} \chi'^2)$}

\subsubsection{Scalar Field}

Since the right-hand side of Eq.~\eqref{scalar-wave-eq} is already proportional to $\alpha/\beta$, we expand the Pontryagin density $R_{\nu\mu \rho \sigma} {}^* R^{\mu\nu\rho\sigma}$ only up to $\mathcal{O}(\alpha'^0)$.
This means that we can substitute the Kerr solution in $R_{\nu\mu \rho \sigma} {}^* R^{\mu\nu\rho\sigma}$ and expand it in powers of $\chi'$ to find
\ba
R_{\nu\mu \rho \sigma} {}^* R^{\mu\nu\rho\sigma} &=& 288 \frac{M^3 \chi \cos \theta}{r^7} \left( 1-\frac{28}{3} \frac{M^2}{r^2} \chi^2 \cos^2 \theta \right) \nn \\
& & + \mathcal{O}(\chi'^5)\,.
\label{pont}
\ea
The Pontryagin density is a parity odd quantity, and hence it can only depend on odd powers of $\chi'$. 
Since we are only interested in a BH solution to $\mathcal{O}(\alpha'^{2} \chi'^2)$, we need not concern ourselves with the second term in Eq.~\eqref{pont}. Therefore, one finds that 
\be 
\vartheta^{(2,1)}=0\,,
\ee
and we only have to consider $\vartheta^{(1,1)}$ [see Eq.~\eqref{theta-linear}] to construct $g_{\mu\nu}^{(2,2)}$. In fact, this shows that $\vartheta^{(n,1)} = 0$ for all even $n$.

\subsubsection{Metric Tensor: Equations}

Consider an expansion of the metric of the form $g_{\mu\nu}=g_{\mu\nu}^{(0,0)}+h_{\mu\nu}$ where $h_{\mu\nu}$ denotes a metric perturbation away from Schwarzschild solution. For us, this metric deformation contains both known terms, such as pure Kerr deformations of Schwarzschild and CS corrections at ${\cal{O}}(\alpha'^{2} \chi')$, as well as unknown terms, such as CS corrections at ${\cal{O}}(\alpha'^{2} \chi'^{2})$, namely
\be
h_{\mu \nu} = \chi' g_{\mu \nu}^{(1,0)} + \chi'{}^{2} g_{\mu \nu}^{(2,0)} + \chi' \alpha'{}^{2} g_{\mu \nu}^{(1,2)} + \chi'{}^{2} \alpha'{}^{2}  g_{\mu \nu}^{(2,2)}\,.
\ee
The Einstein tensor can then be expanded as
\be
G_{\mu\nu} = G_{\mu\nu}^{[0]} + G_{\mu\nu}^{[1]} \left[ h_{\mu\nu} \right] + G_{\mu\nu}^{[2]} \left[ h_{\mu\nu}, h_{\mu\nu} \right] + \mathcal{O}(h^3)\,.
\label{G-exp}
\ee
Here, the superscript in square brackets counts the number of times $h_{\mu\nu}$ appears. Obviously, the first term in Eq.~\eqref{G-exp} vanishes because the Schwarzschild metric satisfies the vacuum Einstein equations. 

With this notation, we can split the ${\cal{O}}(\alpha'^{2} \chi'^{2})$ part of the Einstein tensor $G_{\mu\nu}^{(2,2)}$ into two contributions 
\ba
G_{\mu\nu}^{(2,2)} = G_{\mu\nu}^{[1]}\left[ g_{\mu\nu}^{(2,2)} \right] + G_{\mu\nu}^{[2]} \left[ g_{\mu\nu}^{(1,0)}, g_{\mu\nu}^{(1,2)} \right]\,,
\ea
where the first term is constructed from the unknown functions $g_{\mu\nu}^{(2,2)}$ and its derivatives, while the second term is a known function built from $g_{\mu\nu}^{(1,0)}$ and $g_{\mu\nu}^{(1,2)} $ only. We can then rewrite the field equations at $\mathcal{O}(\alpha'^{2} \chi'^2)$ as
\begin{align}
G_{\mu\nu}^{[1]}\left[ g_{\mu\nu}^{(2,2)} \right]  &=  S_{\mu\nu}^{(2,2)}\,,
\label{field-eq-21}
\end{align}
where we have defined the source term
\begin{align}
S_{\mu\nu}^{(2,2)} \equiv - G_{\mu\nu}^{[2]} \left[ g_{\mu\nu}^{(1,0)}, g_{\mu\nu}^{(1,2)} \right]  -C'_{\mu\nu}{}^{(2,2)}+T_{\mu\nu}'^{\vartheta}{}^{(2,2)}\,.
\end{align}
For convenience, we introduced the reduced C-tensor $C'_{\mu\nu}$ and the reduced stress energy-momentum tensor of the scalar field $T_{\mu\nu}^{\vartheta}{}' $:
\be
C'_{\mu\nu} \equiv (\alpha/\kappa_g) C_{\mu\nu}\,, \quad T_{\mu\nu}'^{\vartheta} \equiv (1/2 \kappa_g) T_{\mu\nu}^{\vartheta}\,.
\ee
The components of $S_{\mu \nu}^{(2,2)}$ can be calculated from Eqs.~\eqref{theta-linear}, \eqref{metric-linear} and the Kerr metric.

The recasted field equations carry a strong resemblance with the equations of BH perturbation theory~\cite{zerilli,sago}. The quantity on the left-hand side of Eq.~\eqref{field-eq-21} can be interpreted as the Einstein tensor constructed from the unknown perturbation $g_{\mu\nu}^{(2,2)}$ in a Schwarzschild background $g_{\mu\nu}^{(0,0)}$. Since the source $S_{\mu \nu}^{(2,2)}$ is an analytic function that can be computed exactly, $g_{\mu\nu}^{(2,2)}$ can be solved for using Schwarzschild BH perturbation theory tools.

Following Refs.~\cite{zerilli,sago}, we first decompose the metric perturbation $g_{\mu\nu}^{(2,2)}$ and the source $S_{\mu\nu}^{(2,2)}$ in tensor spherical harmonics. Since terms of $\mathcal{O}(\alpha'^{2} \chi'^2)$ are parity even, we only consider the metric perturbation in the even-parity sector, which has $7$ independent metric components. Imposing stationarity and axisymmetry reduces the number of independent components to $5$. The latter condition allows us to consider the $m=0$ mode only in the spherical harmonic decomposition. Two gauge degrees of freedom remain, which we fix by adopting the Zerilli gauge. One is then left with $3$ independent degrees of freedom, which allows us to parameterize the metric perturbation via 
\ba
\bm{g}^{(2,2)} &=& \sum_{l} \Big[ f(r) H_{0\ell 0}(r) \bm{a}^{(0)}_{\ell 0} + \frac{1}{f(r)} H_{2\ell 0}(r) \bm{a}_{\ell 0} 
\nn \\  & & + \sqrt{2}K_{\ell 0}(r) \bm{g}_{\ell 0} \Big]\,.
\label{metric-decomp}
\ea
and the source term via
\ba
\bm{S}^{(2,2)} &=& \sum_{l} \Big[ A_{\ell 0}^{(0)} \bm{a}_{\ell 0}^{(0)} + A_{\ell 0} \bm{a}_{\ell 0} + B_{\ell 0} \bm{b}_{\ell 0} \nn \\
& &+G_{\ell 0}^{(s)} \bm{g}_{\ell 0} + F_{\ell 0} \bm{f}_{\ell 0} \Big]\,,
\ea
with $f(r) \equiv 1-2M/r$ and the five tensor spherical harmonics $\bm{a}^{(0)}_{\ell 0}$, $\bm{a}_{\ell 0}$, $\bm{b}_{\ell 0}$, $\bm{g}_{\ell 0}$ and $\bm{f}_{\ell 0}$ defined in Appendix~\ref{app:harmonics}. Notice that boldfaced quantities here refer to rank-2 covariant tensors. The source term coefficients $A_{\ell 0}^{(0)}$, $A_{\ell 0}$, $B_{\ell 0}$, $G_{\ell 0}^{(s)}$ and $F_{\ell 0}$ are non-vanishing only for $\ell =0$ and $\ell =2$, and we provide explicit expressions for them in Appendix~\ref{app:harmonics}.

Using this decomposition, the field equations in Eq.~\eqref{field-eq-21} are no longer coupled, partial differential equations, but they rather become coupled ordinary differential equations for $(H_{0\ell 0},H_{2\ell 0},K_{\ell 0})$, namely~\cite{zerilli,sago}
\bw
\ba
\label{Sago1}
& & f(r)^2 \frac{d^2 K_{\ell 0}}{d r^2} + \frac{1}{r} f(r) \left( 3-\frac{5M}{r} \right) \frac{d K_{\ell 0}}{d r} - \frac{1}{r} f(r)^2 \frac{d H_{2\ell 0}}{d r} 
\nn \\
& &\qquad  - \frac{1}{r^2} f(r) (H_{2\ell 0}-K_{\ell 0}) - \frac{l(l+1)}{2r^2}f(r) (H_{2\ell 0}+K_{\ell 0}) = - A_{\ell 0}^{(0)}\,, 
\\
\label{Sago2}
& & -\frac{r-M}{r^{2} f(r)} \frac{d K_{\ell 0}}{d r} + \frac{1}{r} \frac{d H_{0\ell 0}}{d r} + \frac{1}{r^{2}f(r)} (H_{2\ell 0}-K_{\ell 0}) + \frac{l (l+1)}{2r^{2} f(r)} (K_{\ell 0}-H_{0\ell 0}) = - A_{\ell 0}\,, 
\\
\label{Sago3}
& & f(r) \frac{d}{d r} (H_{0\ell 0}-K_{\ell 0}) + \frac{2M}{r^2} H_{0\ell 0} + \frac{1}{r} \left( 1-\frac{M}{r} \right) (H_{2\ell 0}-H_{0\ell 0}) = \frac{r f(r)}{\sqrt{l(l+1)/2}} B_{\ell 0}\,, 
\\
\label{Sago4}
& & f(r) \frac{d^2 K_{\ell 0}}{d r^2} + \frac{2}{r} \left( 1-\frac{M}{r} \right) \frac{d K_{\ell 0}}{d r} - f(r)\frac{d^2 H_{0\ell 0}}{d r^2} -\frac{1}{r} \left( 1-\frac{M}{r} \right) \frac{d H_{2\ell 0}}{d r} \nn \\
& & \qquad - \frac{r+M}{r^2} \frac{d H_{0\ell 0}}{d r} + \frac{l(l+1)}{2r^2} (H_{0\ell 0}-H_{2\ell 0})=\sqrt{2} G_{\ell 0}^{(s)}\,, 
\\
\label{Sago5}
& & \frac{H_{0\ell 0}-H_{2\ell 0}}{2} = \frac{ r^2 F_{\ell 0}}{\sqrt{l(l+1)(l-1)(l+2)/2}}\,.
\ea
\ew
In Eqs.~\eqref{Sago1},~\eqref{Sago2} and~\eqref{Sago4}, $\ell = 0$ or $2$, while in Eqs.~\eqref{Sago3} and~\eqref{Sago5} $\ell = 2$. We have checked that these equations are identical to those derived from the field equations directly through the use of symbolic manipulation software. 

\subsubsection{Metric Tensor: Solution}
\label{sec:solution}

Before solving these equations, let us exhaust the remaining gauge freedom in the $\ell=0$ mode. We already explained that for modes with $\ell \geq 2$, one is left with $5$ independent variables after imposing stationarity and axisymmetry. Mathematically, these variables are contained in the coefficients of the five spherical tensor harmonics $\bm{a}^{(0)}_{\ell 0}$, $\bm{a}_{\ell 0}$, $\bm{b}_{\ell 0}$, $\bm{g}_{\ell 0}$ and $\bm{f}_{\ell 0}$. We eliminate two of them by imposing the Zerilli gauge, i.e. setting the coefficients of $\bm{b}_{\ell 0}$ and $\bm{f}_{\ell 0}$ to zero. The $\ell =0$ mode, however, is different because from the start it possesses only $3$ independent variables, after imposing stationary and axisymmetry. One of them corresponds to a redefinition of the spherical areal radius, which we eliminate by setting $K_{00}=0$.

Let us now discuss how to solve the differential system in Eqs.~\eqref{Sago1}-\eqref{Sago5}. When we substitute $K_{00}=0$ in Eq.~\eqref{Sago1} with $\ell =0$, we are left with a first-order ordinary differential equation for $H_{200}$. We can solve for $H_{200}$ and then use Eqs~\eqref{Sago2} and~\eqref{Sago4} to find $H_{000}$. With this, we can then obtain the $\ell = 2$ perturbative modes to find the $(H_{0\ell 0}, H_{2\ell 0}, K_{\ell 0})$ functions that we present in Appendix~\ref{app:harmonics} for completeness. Each of these solutions is composed of the sum of a homogeneous and an inhomogeneous solution. The former introduces integration constants chosen by requiring (i) that the metric be asymptotically flat at spatial infinity, e.g.~$H_{0\ell0} \rightarrow 0$ as $r \rightarrow \infty$, and (ii) that the mass and spin angular momentum associated with the new solution is given by $M$ and $M a$, as measured by an observer at spatial infinity. 

The line element to $\mathcal{O}(\alpha'^{2} \chi'^2)$ is then $ds^{2} = ds^{2}_{K} + \delta(ds^{2})_{\CS}$, where
\be
\delta(ds^{2})_{\CS}= 2 g_{t\phi}^\CS dtd\phi + g_{tt}^\CS dt^2 + g_{rr}^\CS dr^2 + g_{\theta \theta}^\CS d\theta^2 + g_{\phi\phi}^\CS d\phi^2
\label{the-solution}
\ee
\bw
with
\ba
g_{tt}^\CS &=& \zeta \chi^2 \frac{M^3}{r^3} \Bigg[  \frac{201}{1792} \left( 1+\frac{M}{r} +\frac{4474}{4221} \frac{M^2}{r^2} -\frac{2060}{469} \frac{M^3}{r^3} +\frac{1500}{469} \frac{M^4}{r^4} - \frac{2140}{201} \frac{M^5}{r^5}  + \frac{9256}{201} \frac{M^6}{r^6} - \frac{5376}{67} \frac{M^7}{r^7}  \right) (3\cos^2 \theta -1)  \nn \\
& & - \frac{5}{384} \frac{M^2}{r^2} \left( 1 + 100 \frac{M}{r} + 194\frac{M^2}{r^2} + \frac{2220}{7} \frac{M^3}{r^3} - \frac{1512}{5} \frac{M^4}{r^4} \right) \Bigg]     + \mathcal{O}(\alpha'^2 \chi'^4)\,, \nn \\
\\
g_{rr}^\CS &=&   \zeta \chi^2 \frac{M^3}{r^3 f(r)^2} \Bigg[  \frac{201}{1792}  f(r) \left( 1+ \frac{1459}{603} \frac{M}{r} +\frac{20000}{4221} \frac{M^2}{r^2} +\frac{51580}{1407} \frac{M^3}{r^3} -\frac{7580}{201} \frac{M^4}{r^4} \right. \nn \\ 
& & \left. - \frac{22492}{201} \frac{M^5}{r^5}  - \frac{40320}{67} \frac{M^6}{r^6} \right) (3 \cos^2 \theta -1) \\  \nn 
& & - \frac{25}{384} \frac{M}{r} \left( 1 + 3\frac{M}{r} + \frac{322}{5} \frac{M^2}{r^2} + \frac{198}{5} \frac{M^3}{r^3} + \frac{6276}{175} \frac{M^4}{r^4} - \frac{17496}{25} \frac{M^5}{r^5}   \right) \Bigg]     + \mathcal{O}(\alpha'^2 \chi'^4)\,,
\\
g_{\theta\theta}^\CS &=& \frac{201}{1792} \zeta \chi^2 M^2 \frac{M}{r} \left( 1 + \frac{1420}{603} \frac{M}{r} + \frac{18908}{4221} \frac{M^2}{r^2} + \frac{1480}{603} \frac{M^3}{r^3} + \frac{22460}{1407} \frac{M^4}{r^4} + \frac{3848}{201} \frac{M^5}{r^5} + \frac{5376}{67} \frac{M^6}{r^6} \right) (3 \cos^2 \theta -1) \nn \\
& & + \mathcal{O}(\alpha'^2 \chi'^4)\,, 
\\
g_{\phi\phi}^\CS &=&  \sin^2 \theta g_{\theta\theta}^\CS + \mathcal{O}(\alpha'^2 \chi'^4)
\ea
\ew
and $g_{t\phi}^\CS$ given in Eq.~\eqref{YP-sol}. 
We have checked explicitly that the solution above satisfies the field equations [Eq.~\eqref{field-eq-21}] to $\mathcal{O}(\alpha'^{2} \chi'^2)$ with symbolic manipulation software.

The choice of homogeneous integration constants depend on how we choose to define the mass $M$ and the reduced spin angular momentum $a$. The most natural choice is to define these quantities as measured by an observer at infinity, which then leads to the metric displayed above. With these definitions, the angular velocity and area of the event horizon are modified to 
\ba
\Omega_{\hor} & \equiv & -\frac{g_{tt}}{g_{t\phi}} \Big|_{r=r_{\hor}} = \Omega_{\horK} \left( 1-\frac{709}{7168} \zeta \right)\,, \\
A_{\hor} & \equiv & 2\pi \int^{\pi}_{0} \sqrt{g_{\theta \theta}  g_{\phi \phi}}|_{r=r_{\hor}} d\theta \nn \\
& = & A_{\horK}\left( 1-\frac{915}{28672}\zeta \chi^2 \right)\,,
\ea
where $r_{\hor}$ is the location of the horizon, which will be discussed in the next section (Eq.~\eqref{hor}), and where  $\Omega_{\horK} = a/(r_{\horK}^2+a^2)$ and $A_{\horK} = 16\pi M^2 (1-\chi^2/4) + \mathcal{O}(\chi^4)$, with $r_{\horK}$ the horizon radius for the Kerr metric: $r_{\horK} = M+\sqrt{M^2 - a^2}$. One can physically interpret this result by thinking of the BH metric found here as representing a BH surrounded by a scalar field ``cloud'' with a dipole density structure. As such, the modified BH horizon area and its angular velocity will be modified from what one would expect for a Kerr BH in vacuum GR due to the presence of the scalar field.  

One could of course insist on defining the mass and reduced angular momentum such that the horizon structure of the modified BH remains identical to that of the Kerr metric. This would require a renormalization of $M$ and $a$ via 
\ba
\label{mass}
\widetilde{M} & \equiv & M \left( 1-\frac{2333}{57344} \zeta \chi^2 \right)\,, \\
\label{spin}
\tilde{a} & \equiv & a \left( 1-\frac{709}{7168} \zeta \right)\,,
\ea
which then leads to $\Omega_{H}=\tilde{\Omega}_{\horK}$ and $A_{\hor}=\tilde{A}_{\horK}$ where a quantity with a tilde means that $M$ and $a$ appearing in that quantity are replaced by $\widetilde{M}$ and $\tilde{a}$. One can think of $\widetilde{M}$ and $\tilde{a}$ as a ``bare''  BH mass and reduced angular momentum. The asymptotically defined $M$ and $a$ are larger than the bare ones due to the cloud of scalar field outside the BH. 

Another quasi-local measure of mass $M_{\Komar}$ and reduced spin angular momentum $a_{\Komar}$ can be defined by the Komar integrals~\cite{toolkit}. For the modified BH metric in Eq.~\eqref{the-solution} one finds
\ba
M_{\Komar} &\equiv & \frac{1}{2} \int_0^{\pi} \left( t^{\alpha ;\beta} n_{[\alpha} r_{\beta ]} \sqrt{g_{\theta \theta} g_{\phi \phi}} \right) |_{r=r_\hor} d\theta \nn \\
&=& M \left( 1-\frac{1727}{14336} \zeta \chi^2 \right)\,, \\
a_{\Komar} &\equiv & -\frac{1}{4M} \int_0^{\pi} \left( \phi^{\alpha ;\beta} n_{[\alpha} r_{\beta ]} \sqrt{g_{\theta \theta} g_{\phi \phi}} \right) |_{r=r_\hor} d\theta  \nn \\
&=& a\left( 1-\frac{29}{128}\zeta \right)\,,
\ea
where $t^\alpha \partial_\alpha=-\partial /\partial t$ and $\phi^\alpha \partial_\alpha=\partial /\partial \phi$ are timelike and spacelike Killing vectors, while $n_\alpha$ and $r_\alpha$ are the unit covariant vectors normal to the $t=\mrm{const.}$ and $r=\mrm{const.}$ hypersurfaces, respectively. Notice that the above quantities are calculated at $r=r_\hor$, which are different from the so-called Komar mass and angular momentum which is defined over a 2-sphere at spatial infinity~\cite{toolkit}. As before, the above quasilocal quantities are smaller than those defined at spatial infinity due to the presence of the scalar field cloud close to the modified BH. One could re-express the metric in terms of $(M_{\Komar},a_{\Komar})$ or $(\widetilde{M},\tilde{a})$, but we choose not to do so as these quantities are not the masses and angular momenta that an observer at infinity would measure. 

One might be worried that the solution presented in Eq.~\eqref{the-solution} diverges at the unperturbed Schwarzschild horizon. This is simply a spurious divergence that arises due to the slow-rotation expansion, even for the Kerr metric. For example, if one takes the $(r,r)$ component of the Kerr metric in BL coordinates and expands it in $\chi \ll 1$, one finds terms to leading order that scale as $(r - 2 M)^{-1}$, which diverges at the Schwarzschild horizon. The unexpanded Kerr metric, however, can have a horizon located inside $2M$ for prograde spins. One can then be faced with the unpleasant situation of the slow-rotation expansion of $g_{rr}$ diverging outside the true event horizon due to the slow-rotation approximation.

For practical reasons, one might wish to eliminate this feature through resummation. By the latter, we mean a modification of certain terms in the metric that naively diverge at the Schwarzschild or unperturbed Kerr horizon, such that 
\begin{enumerate}
\item when the resummed metric is expanded in $\chi \ll1$, it becomes identical to the old metric to a given order in $\chi$, i.e. $\rm{\hat{E}}[g_{\mu \nu}^{\rm resum}] = \rm{\hat{E}}[g_{\mu \nu}]$. 
\item all components of the resummed metric $g_{\mu \nu}^{\rm resum}$ remain finite everywhere outside the dynamical CS modified horizon. 
\end{enumerate}
where $g_{\mu \nu}^{\rm resum}$ is the resummed metric, $g_{\mu \nu}$ is the metric of Eq.~\eqref{the-solution} and the $\hat{\rm{E}}[\cdot]$ operator stands for expansion in $\chi \ll 1$. 

In principle, there is an infinite number of ways in which one can resum the metric. One way is to replace $\Delta \to \Delta'$ in $g_{rr,{\K}}$ [i.e.~in the $(r,r)$ component of the Kerr metric in Eq.~\eqref{Kerr-metric}] and $f(r) \to f'(r)$ in $\delta(ds^{2}_{\CS})$ of Eq.~\eqref{the-solution}, where we have defined 
\begin{align}
\Delta' &\equiv \Delta + \frac{915}{14336} \zeta M^{2} \chi^{2}\,,
\\
f'(r) &\equiv 1 - \frac{r_{H}}{r}\,. 
\end{align}
To retain the asymptotic behavior in the $\chi \ll 1$ limit, one then needs to add the following counterterm in the $(r,r)$ component of the metric (induced by the $\Delta'$ modification to the Kerr metric):
\begin{equation}
\delta g_{rr} = \frac{915}{14336} \frac{\zeta M^{2} \chi^{2}}{r^2 f'(r)^{2}}\,.
\end{equation}
With these changes, the resummed metric is 
\be
g_{\mu \nu}^{\rm resum} = g_{\mu \nu,{\K}}[\Delta'] +  g_{\mu \nu}^{\CS}[f'] + \delta g_{rr} \delta_{\mu}^{r} \delta_{\nu}^{r} \,,
\ee
will satisfy the conditions enumerated above. 

\section{Properties of the solution}
\label{sec:properties}

In this section, we discuss various properties of the new solution.
We begin by finding the corrected locations of the horizon and ergosphere.
We continue with a calculation of the corrected quadrupole moment of the spacetime.
We then determine the Petrov type of the new solution.

\subsection{Singularity, Horizon, and Ergosphere}

The spacetime in Eq.~\eqref{the-solution} contains a true singularity at $r=0$. This can be verified by computing the Kretchmann invariant $R_{\mu\nu\rho\sigma} R^{\mu\nu\rho\sigma}$, which diverges at $r=0$. Indeed, this quantity is identical to that found in~\cite{yunespretorius} to ${\cal{O}}(\alpha'^2 \chi')$. We do not present the ${\cal{O}}(\alpha'^2 \chi'^{2})$ term here, as this cannot cure the $r = 0$ divergence.

The location of the event horizon can be found by solving the equation $g_{tt} g_{\phi\phi} - g_{t\phi}^2 =0$ for $r$~\cite{toolkit}.
We find
\be
r_{\hor}=r_{\horK} - \frac{915}{28672} \zeta M \chi^2 + \mathcal{O}(\alpha'^2 \chi'^3)\,,
\label{hor}
\ee
The horizon radius decreases relative to the Kerr horizon radius, but of course, the shift of the horizon location depends on how one renormalizes the mass and spin and also on the choice of radial coordinate. 

The location of the ergosphere can be found by solving the equation $g_{tt}=0$ for $r$. We find
\ba
r_\mrm{ergo} &=& r_\mrm{ergo,{\K}} -\frac{915}{28672} \zeta M \chi^2 \left( 1+\frac{2836}{915} \sin^2 \theta \right) \nn \\
& & + \mathcal{O}(\alpha'^2 \chi'^3)
\ea
with the ergosphere of the Kerr solution given by $r_\mrm{ergo,{\K}} = M + \sqrt{M^2 - a^2 \cos^2 \theta} \approx r_{\horK} [1+(\chi/4)\sin^2 \theta]$. 

\subsection{Lorentz Signature}
We show the Lorentzian signature of the metric is preserved outside the horizon, provided the coupling constant is small. Otherwise, our perturbative construction of solution will not be justified well.
By denoting the determinant of the metric component as $g$ and the one for the Kerr as $g_{\K} \equiv -r^2 \sin^2\theta (r^2+a^2 \cos^2\theta)+\mathcal{O}(\chi'^3)$, $g/g_{\K}$ is given by
\bw
\begin{align}
\frac{g}{g_{\K}} &= 1+\frac{4211}{6272} \zeta \frac{M^3}{r^3} \chi^2 \left( 1+\frac{415}{201} \frac{M}{r} + \frac{14008}{4221} \frac{M^2}{r^2} + \frac{3200}{201} \frac{M^3}{r^3} - \frac{14780}{1407}\frac{M^4}{r^4} - \frac{8108}{201} \frac{M^5}{r^5} - \frac{16128}{67} \frac{M^6}{r^6} \right) \cos^2\theta \nn \\
&-\frac{201}{896} \zeta \frac{M^3}{r^3} \chi^2 \left( 1+\frac{1420}{603} \frac{M}{r} + \frac{19888}{4221} \frac{M^2}{r^2} + \frac{6350}{201} \frac{M^3}{r^3} + \frac{40100}{1407}\frac{M^4}{r^4} + \frac{8524}{201} \frac{M^5}{r^5} - \frac{16128}{67} \frac{M^6}{r^6} \right) + {\cal{O}}(\alpha'^2 \chi'^{3})\,. 
\end{align}
\ew
Notice that $g/g_{\K}$ does not diverge at $r = 2M$, just like the resummed metric. This is because the determinant of the metric is given by $g = g_{rr} g_{\theta \theta} (g_{tt} g_{\phi\phi}-g_{t \phi}^2)$, and while $g_{rr} \propto f(r)^{-2}$, the quantity $(g_{tt} g_{\phi\phi}-g_{t \phi}^2) \propto \Delta^{2} \sim f(r)^{2}$ and thus $g$ is finite at $r = 2M$. Since the correction terms fall off rapidly as $r\rightarrow \infty$, it is important to look at the signature of $g/g_{\K}$ at the horizon, which is given by
\be
\frac{g}{g_{\K}} = 1 - \frac{74849}{401408} \zeta \chi^2  \left( 1+ \frac{27901}{74849} \cos^2 \theta \right) + {\cal{O}}(\alpha'^2 \chi'^{3})\,.
\ee
The correction terms above are negative for any $\theta$.
The magnitude of the correction to $g/g_{\K}$ becomes the largest at the poles and at the equatorial plane, respectively.
One can see that within the small-coupling and slow-rotation regime, the signature flip does not take place.

\subsection{Closed Timelike Curves}
The new BH solution contains no closed timelike curves (CTCs) outside the horizon. If they existed, these curves could be found by solving for the region where $g_{\phi \phi}>0$. The explicit forms of $g_{\phi\phi}$ at the horizon is given by
\begin{align}
g_{\phi\phi} &= 4\sin^2 \theta M^2 \Bigg[ 1  - \frac{1}{4} \chi^2 \cos^2 \theta  \nn \\
&-\frac{12283}{100352} \zeta \chi^2 \left( 1 - \frac{54483}{24566} \cos^2 \theta \right) \Bigg]+ {\cal{O}}(\alpha'^2 \chi'^{3}) \,, 
\label{gphph}
\end{align}
where we note that $g_{\phi\phi}$ vanishes at the poles. The correction terms are positive in the polar region and negative in the equatorial region.
Eq.~\eqref{gphph} clearly shows that small perturbation due to CS coupling does not change the causal structure of spacetime. 

\subsection{Multipolar Structure}

Since $M$ and $a$ are asymptotic quantities, the first non-vanishing correction to the spacetime's multipolar structure on $g_{\mu\nu}$ appears in the mass quadrupole moment. Following Thorne~\cite{thorne-MM}, we
can read off the multipole moments by transforming the metric from BL-type coordinates  to so-called asymptotically Cartesian and mass centered (ACMC) coordinates (the coordinates where the multipole moments are defined in a spacetime region asymptotically far from the source). In order to determine the quadrupole moment, we need to transform to ACMC coordinates such that $g_{tt}$ and $g_{ij}$ at $\mathcal{O}(r^{-2})$ do not contain any angle dependence. In these coordinates, the metric component $g_{tt}$ for a stationary and axisymmetric spacetime can be expressed as
\be
g_{tt}=-1+\frac{2M}{r} + \frac{\sqrt{3}}{2} \frac{1}{r^3} \left[ I_{20} Y_{20} + (l=0 \ \mrm{pole})  \right] + \mathcal{O} \left(\frac{1}{r^4}  \right)\,.
\label{Eq:Thorne}
\ee
Here, $Y_{20}$ is the $(\ell,m)=(2,0)$ spherical harmonic, and $I_{20}$ corresponds to ($m=0$) quadrupole moment.

Let us first extract the quadrupole moment of a Kerr BH. By choosing the {\emph{flat-spacetime normalized}} basis
\be
\bm{e}_0=\partial_t, \quad \bm{e}_r = \partial_r, \quad \bm{e}_\theta = r^{-1} \partial_\theta, \quad \bm{e}_\phi = (r \sin \theta)^{-1} \partial_\phi\,,
\ee
associated with BL coordinates, the Kerr metric can be re-expressed as
\ba
\bar{g}_{tt,{\K}} &=& -1 + \frac{2M}{r} - \frac{2Ma^2 \cos^2 \theta}{r^3} + \mathcal{O}\left( \frac{1}{r^5} \right)\,,  \\
\bar{g}_{t\phi,{\K}} &=& - \frac{2Ma\sin\theta}{r^2} + \frac{2Ma^3 \sin\theta \cos^2\theta}{r^4} + \mathcal{O}\left( \frac{1}{r^6} \right)\,, \nn \\
\\
\bar{g}_{rr,{\K}} &=& 1+\frac{2M}{r} + \frac{4M^2-a^2 \sin^2 \theta}{r^2} \nn \\
& & + \frac{8M^3-2Ma^2(2-\cos^2\theta)}{r^3} + \mathcal{O}\left( \frac{1}{r^4} \right)\,,  \\
\bar{g}_{\theta\theta,{\K}} &=& 1+ \frac{a^2 \cos^2 \theta}{r^2}\,, \\
\bar{g}_{\phi\phi,{\K}} &=& 1+\frac{a^2}{r^2}+\frac{2Ma^2 \sin^2\theta}{r^3} +\mathcal{O}\left( \frac{1}{r^5} \right)\,.  
\ea
In order to read off the quadrupole moment, we need to transform to appropriate ACMC coordinates $(t',r',\theta',\phi')$ such that $-a^2 \sin^2\theta/r^2$ in $\bar{g}_{rr,{\K}}$ and $\bar{g}_{\theta\theta,{\K}}$ disappears. This can be realized by the transformation
\ba
t&=&t'\,, \\
r&=& r' + \frac{a^2\cos^2\theta'}{2r'}\,, \\
\theta &=& \theta' - \frac{a^2 \cos\theta' \sin\theta'}{2 r'^2}\,, \\
\phi &=& \phi'\,.
\label{transfrom}
\ea
In this ACMC coordinates, $g'_{tt,{\K}}$ becomes
\be
g'_{tt,{\K}} = -1 + \frac{2M}{r'} - \frac{3Ma^2 \cos^2 \theta'}{r'^3} + \mathcal{O}\left( \frac{1}{r'^5} \right)\,.
\ee
Therefore, by comparing this with Eq.~\eqref{Eq:Thorne}, one can read off the quadrupole moment as~\cite{thorne-MM} 
\be
I_{20,{\K}} = -8\sqrt{\frac{\pi}{15}} Ma^2\,.
\label{Kerr-quadrupole}
\ee

Let us now follow the above procedure to determine the quadrupole moment of the new BH solution.
Since the correction in the metric is already to $\mathcal{O}(\alpha'^2 \chi'^2)$, it is not affected by the above coordinate transformation.
The quadrupole moment in the new solution can then be read off as
\be
I_{20} = I_{20,{\K}} \left( 1-\frac{201}{1792} \zeta \right)\,.
\ee
Notice that this correction vanishes to linear order in the spin; the linear-order in spin terms correct the multipolar structure of the spacetime at much higher multipole order~\cite{sopuertayunes}. 

Geroch and Hansen~\cite{geroch,hansen} proposed a slightly different definition of multipole moments, which for example leads to a $I_{20,{\K}}$ that differs from Eq.~\eqref{Kerr-quadrupole} by a factor $8\sqrt{\pi/15}$. However, this difference is just a matter of convention. One should realize, of course, that the quadrupole moment itself is not a directly observable quantity. Modifications to the BH multipolar structure, however, do imprint on the motion of massive and massless bodies. Corrections to the gravitational radiation induced by this modified motion is indeed observable.  

\subsection{Petrov Type}
\label{app:petrov}

A generic spacetime can be classified into Petrov types by counting the number of distinct principal null directions (PNDs) $k^\mu$ of the Weyl tensor $C_{\mu\nu\rho\sigma}$~\cite{stephani,campanelli}, where $k^\mu$ satisfies
\be
k^\nu k^\rho k_{[\tau}C_{\mu ]\nu\rho [\sigma} k_{\chi ]}=0\,.
\ee
This is equivalent to finding the number of distinct PNDs $l^\mu$ that make one of the Weyl scalars $\Psi_0 =0$, which reduces to counting the number of distinct roots of the following equation for $b$~\cite{stephani}:
\be
\Psi_0 + 4b \Psi_1 + 6b^2 \Psi_2 + 4 b^3 \Psi_3 + b^4 \Psi_4 =0\,.
\label{quartic}
\ee
Here, $\Psi_0$,...$\Psi_4$ are five complex Weyl scalars in an arbitrary tetrad with the restriction $\Psi_4 \neq 0$.

If Eq~\eqref{quartic} contains at least one degenerate root, the spacetime is said to be {\emph{algebraically special}} and the following relation holds:
\be
I^3 = 27 J^2\,.
\label{S}
\ee
Here, the quadratic and cubic Weyl quantities $I$ and $J$ are defined by~\cite{stephani}
\ba
\label{I-eq}
I &\equiv& \frac{1}{2} \tilde{C}_{\alpha\beta\gamma\delta}\tilde{C}^{\alpha\beta\gamma\delta} \nn \\
& =& 3 \Psi_2^2-4\Psi_1\Psi_3 + \Psi_4 \Psi_0\,, \\ 
\label{J-eq}
J &\equiv & -\frac{1}{6}  \tilde{C}_{\alpha\beta\gamma\delta} \tilde{C}^{\gamma \delta}{}_{\mu\nu}\tilde{C}^{\mu\nu\alpha\beta} \nn \\
&=& -\Psi_2^3 + 2 \Psi_1 \Psi_3 \Psi_2 + \Psi_0 \Psi_4 \Psi_2 - \Psi_4 \Psi_1^2 - \Psi_0 \Psi_3^2 \nn \\
\ea
with
\be
\tilde{C}_{\alpha\beta\gamma\delta} \equiv \frac{1}{4} \left( C_{\alpha\beta\gamma\delta} + \frac{i}{2}\epsilon_{\alpha\beta\mu\nu} C^{\mu\nu}{}_{\gamma\delta} \right)\,.
\ee
If Eq.~\eqref{S} is not satisfied, the spacetime is of Petrov type I.
The Kerr BH and the slowly-rotating BH in dynamical CS gravity to $\mathcal{O}(\alpha'^2 \chi')$ is known to be of Petrov type D, which means that Eq.~\eqref{quartic} has double degenerate roots.
In type D spacetimes, not only Eq.~\eqref{S} holds but there are additional conditions that need to be satisfied:
\be
K=0, \quad N-9L^2=0\,,
\label{conditionD}
\ee
where $K$, $L$ and $N$ are defined as
\ba
K & \equiv & \Psi_1 \Psi_4^2 - 3 \Psi_4 \Psi_3 \Psi_2 + 2 \Psi_3^3\,, \\
\label{L}
L & \equiv & \Psi_2 \Psi_4 - \Psi_3^2\,, \\
\label{N}
N & \equiv & \Psi_4^2 I - 3 L^2 \nn \\
&=& \Psi_4^3 \Psi_0 - 4 \Psi_4^2 \Psi_1 \Psi_3 + 6 \Psi_4 \Psi_2 \Psi_3^2 - 3 \Psi_3^4\,.
\ea

Equation~\eqref{S} determines whether a spacetime is algebraically special, but when the spacetime is an approximate solution, to what order in perturbation theory should this equation be calculated? Let us first concentrate on the BH metric in dynamical CS gravity with only the odd-parity terms of ${\cal{O}}(\alpha'^{2} \chi')$. One can construct a null tetrad that is a deformation away from the Kerr principal null tetrad, such that $\Psi_{2} = {\cal{O}}(1)$, while $\Psi_{1}$ and $\Psi_{3}$ are of ${\cal{O}}(\alpha'^{2} \chi')$. The remaining Newman-Penrose scalars, $\Psi_{0}$ and $\Psi_{4}$, would vanish to this order. One then sees that the first term in Eqs.~\eqref{I-eq} and~\eqref{J-eq} is of ${\cal{O}}(1)$, the second is of ${\cal{O}}(\alpha'^{4} \chi'^{2})$ and the others vanish to this order. Therefore, Eq.~\eqref{S} is trivially satisfied to ${\cal{O}}(\alpha'^2 \chi')$, while the first non-trivial dynamical CS corrections enters at ${\cal{O}}(\alpha'^{4} \chi'^{2})$. Similarly, if one were studying the new BH metric found in this paper, i.e.~including terms of ${\cal{O}}(\alpha'^{2} \chi'^{2})$, then one would have to consider Eq.~\eqref{S} to ${\cal{O}}(\alpha'^{4} \chi'^{4})$. Obviously, if Eq.~\eqref{S} is not satisfied at ${\cal{O}}(\alpha'^{4} \chi'^{2})$, then one does not need to consider the higher order terms. Notice also that the $I$ and $J$ quantities are invariant, and one could have chosen another tetrad, but the arguments presented above would still hold. 

Now that the order to which terms must be expanded is clear, let us focus again on the BH metric in dynamical CS gravity with only the odd-parity terms of ${\cal{O}}(\alpha'^{2} \chi')$. Sopuerta and Yunes~\cite{sopuertayunes} claimed that this metric is of Petrov type D. We have verified this claim as follows. First, we showed that Eq.~\eqref{S} is satisfied to ${\cal{O}}(\alpha'^{4} \chi'^{2})$. Then, we showed that the relations in Eq.~\eqref{conditionD} are also satisfied to ${\cal{O}}(\alpha'^{4} \chi'^{2})$. Although this is sufficient to claim that the metric to this order is of Petrov type D, we also verified explicitly that Eq.~\eqref{quartic} has double degenerate roots. This then implies that one can rotate the null tetrad to a principal one (of the dynamical CS metric), where $\Psi_{2}$ is non-vanishing and contains Kerr terms [of ${\cal{O}}(\alpha'^{0})$ but with spin corrections], as well as terms of ${\cal{O}}(\alpha'^{2} \chi')$. All other Newman-Penrose scalars vanish at this order, i.e.~they are at least of ${\cal{O}}(\alpha'^{2} \chi'^{2})$.

Let us now focus on the new BH metric in dynamical CS gravity, which includes terms of ${\cal{O}}(\alpha'^{2} \chi'^{2})$. Picking the principal null tetrad of the BH metric in dynamical CS gravity with only the odd-parity terms included, one can show that Eq.~\eqref{S} is not satisfied to ${\cal{O}}(\alpha'^{4} \chi'^{4})$. One might worry that to make this statement precise, one would have to account for terms of ${\cal{O}}(\alpha'^{4} \chi')$ in the gravitomagnetic sector of the dynamical CS metric. These terms, however, would modify Eq.~\eqref{S} at ${\cal{O}}(\alpha'^{6} \chi'^{2})$, and thus they can be neglected. Since Eq.~\eqref{S} is not satisfied to ${\cal{O}}(\alpha'^{4} \chi'^{4})$, the new metric presented in this paper breaks symmetries that the odd-parity BH metric used to have. This suggests the exact BH solution should be of Petrov type I.

\section{Geodesic Motion and Separability, Binding Energy and Kepler's Law}
\label{sec:geodesic}

In this section, we discuss the separability of the geodesic equations in the modified metric and find the binding energy, Kepler's Law and the innermost stable circular orbit (ISCO).

\subsection{Geodesic Motion of Test-Particles and Separability}
\label{sec:separability}

Consider the motion of non-spinning test particles in the new BH solution. We concentrate here on non-spinning objects, as otherwise we would have to introduce an additional scalar dipole-dipole interaction~\cite{quadratic}, which will be investigated elsewhere~\cite{kent-CSGW}. 

One of the most interesting properties of the Kerr metric is that the geodesic equations are Liouville integrable~\cite{bardeen}. This leads to the existence of four constants of motion or invariants (quantities that Poisson commute with the Hamiltonian): the mass, energy, angular momentum and the Carter constant. The existence of this last quantity, found by Carter~\cite{carter}, as well as the use of the proper coordinate system, is crucial in showing that the Hamilton-Jacobi equations are separable. In GR, this is related to the Kerr solution being of Petrov type D~\cite{walkerpenrose1970}, i.e.~its associated Weyl tensor possesses double degenerate principal null directions. As shown in Sec.~\ref{app:petrov}, the new solution derived in this paper is of Petrov type I and there is no guarantee that it possesses a Carter-like constant.

The Carter constant is associated with the existence of a second-rank Killing tensor $\xi_{\alpha\beta}$, which in GR and in BL coordinates is defined by 
\be
\xi_{\alpha \beta} = \Delta k_{(\alpha}l_{\beta )}+r^2 g_{\alpha\beta}
\ee
with two null vectors $k^\alpha$ and $l^\alpha$. The odd-parity BH solution in dynamical CS gravity of~\cite{sopuertayunes} does possess a Killing tensor $\bar{\xi}_{\alpha\beta}$ at $\mathcal{O}(\alpha'^{2} \chi')$, where the null vectors $\bar{k}^\alpha$ and $\bar{l}^\alpha$ are given by
\ba
\bar{k}^\alpha \partial_\alpha & \equiv & \frac{r^2+a^2}{\Delta} \partial_t + \partial_r + \left(\frac{a}{\Delta} -\delta g_{\phi}^\mrm{CS} \right) \partial_\phi\,, \\
 \bar{l}^\alpha \partial_\alpha & \equiv & \frac{r^2+a^2}{\Delta} \partial_t - \partial_r + \left(\frac{a}{\Delta} -\delta g_{\phi}^\mrm{CS} \right) \partial_\phi
\ea
and 
\be
\delta g_{\phi}^{\CS} \equiv \frac{5}{8}\zeta \frac{\chi}{M} \frac{M^6}{r^6 f(r)} \left( 1 + \frac{12}{7} \frac{M}{r} + \frac{27}{10} \frac{M^2}{r^2} \right)\,.
\ee
Moreover, one can show that the two null vectors $k^\alpha$ and $l^\alpha$ are also principal null directions of the spacetime to $\mathcal{O}(\alpha'^{2} \chi')$~\cite{sopuertayunes}.

Let us now study whether a non-trivial second-order Killing tensor continues to exist to ${\cal{O}}(\alpha'^{2} \chi'^{2})$, i.e. we look for a correction $\delta \xi_{\alpha\beta}$ to the Killing tensor
\be
\xi_{\alpha\beta} = \bar{\xi}_{\alpha\beta} + \alpha'^2 \chi'^{2} \delta \xi_{\alpha\beta} + {\cal{O}}(\alpha'^2 \chi'^{3})\,,
\label{xi}
\ee
that satisfies the Killing equation $\nabla_{(\gamma} \xi_{\alpha \beta)} = 0$. If a conserved quantity contains both even and odd parts under the simultaneous reflection $t \rightarrow -t$ and $\phi \rightarrow -\phi$, they should be separately conserved. The new BH metric is symmetric under this simultaneous reflection, and hence any geodesic remains geodesic under this transformation. This means that if we consider a quantity $\xi_{\alpha \beta} u^{\alpha} u^{\beta}$ (here, $u^{\alpha}$ is a four velocity vector), the only non-vanishing components allowed are those even in reflection, i.e.~$(t,t)$, $(t,\phi)$, $(\phi,\phi)$, $(r,r)$, $(r,\theta)$ and $(\theta,\theta)$. Without loss of generality, these six components can be parametrized through six free functions, $A(r,\theta)$, $B(r,\theta)$, $C(r,\theta)$, $D(r,\theta)$, $E(r,\theta)$ and $\delta\xi_{r\theta}(r,\theta)$, through the following ansatz:
\ba
\label{deltaxi}
\delta\xi_{\alpha\beta} &\equiv & A(r,\theta) t_\alpha t_\beta + B(r,\theta) t_\alpha \phi_\beta + C(r,\theta) \phi_\alpha \phi_\beta \nn \\
& & + D(r,\theta) g_{\alpha\beta} + E(r,\theta) \bar{\xi}_{\alpha\beta} + \delta\xi_{r\theta}(r,\theta) r_\alpha \theta_\beta\,,  \nn \\
\ea
where $\theta^\alpha \partial_\alpha \equiv \partial / \partial \theta$. 

From the symmetry arguments described above, the Killing equations contain only 10 independent components. The five functions $\mathcal{F} \equiv (A,B,C,D,E)$ appear only in the form $\partial_r \mathcal{F}$ or $\partial_\theta \mathcal{F}$. Thus, we can solve the 10 Killing equations for the 10 functions $\partial_r \mathcal{F}$ and $\partial_\theta \mathcal{F}$ in terms of $\delta\xi_{r\theta}(r,\theta)$. We have found, however, that the consistency relation $\partial_\theta (\partial_r B) = \partial_r (\partial_\theta B)$ does not hold for any $\delta\xi_{\alpha\beta}$ to $\mathcal{O}(\alpha'^2 \chi'^2)$. Since our ansatz is sufficiently generic, this strongly indicates that there does not exist any non-trivial  second-rank Killing tensor in the new BH solution. 

By applying Theorem 1 in Benenti and Francaviglia~\cite{benenti}, the non-existence of a non-trivial second-rank Killing tensor is enough to claim that the 4 dimensional manifold does not admit a \textit{separability structure}. This means that there does not exist any spacetime coordinate transformation that leads to the Hamilton-Jacobi equation being separable.  Since this is an important point, we have verified it in two additional ways: (i) by performing a Levi-Civita test~\cite{levi-civita,yasui} and (ii) by trying to map the new solution to the most generic spacetime that admits a separability structure~\cite{benenti} [see Appendix~\ref{app:Killing} for more details]. In all cases, it is clear that the new metric does not admit such a structure.

Up until now, we showed in various ways that a non-trivial second-rank Killing tensor that is a {\emph{perturbation}} of the Killing tensor found in~\cite{sopuertayunes} does not exist. One might wonder whether there is a {\emph{completely new}} second-rank Killing tensor that is not a perturbation of that in~\cite{sopuertayunes}. If this exists, 
one must be able to find it by setting $\bar{\xi}_{\alpha \beta}=0$ in Eq.~\eqref{xi}. By imposing the same ansatz of Eq.~\eqref{deltaxi}, we can again solve for $\partial_r \mathcal{F}$ and $\partial_\theta \mathcal{F}$. This time, $\partial_\theta (\partial_r B) = \partial_r (\partial_\theta B)$ is trivially satisfied. However, one finds that the only solution of $\delta \xi_{r \theta}$ that satisfies $\partial_\theta (\partial_r \mathcal{F}) = \partial_r (\partial_\theta \mathcal{F})$ is $\delta \xi_{r \theta}=0$, leading to $\delta \xi_{\alpha \beta}=0$. This proves that a completely new non-trivial 2nd-rank Killing tensor at $\mathcal{O}(\alpha'^2 \chi'^2)$ does not exist. The non-existence of this tensor can be proved in a different manner. If it exists, the leading contribution should start at $\mathcal{O}(\alpha'^2 \chi'^2)$. However, the only possible form of this leading term would be $\alpha'^2 \chi'^2 \bar{\xi}^{\K}_{\alpha \beta}$ where $\bar{\xi}^{\K}_{\alpha \beta}$ is the 2nd-rank Killing tensor of the Kerr spacetime. This is because the completely new Killing tensor divided by $\alpha'^2 \chi'^2$ should also be a Killing tensor whose leading term should satisfy the GR Killing equations. Since $\alpha'^2 \chi'^2 \bar{\xi}^{\K}_{\alpha \beta}$ does not satisfy the Killing equations at $\mathcal{O}(\alpha'^2 \chi'^2)$, we conclude that the completely new 2nd-rank Killing tensor cannot exist. Of course, we cannot rule out the possibility of the existence of the completely new Killing tensor once the small coupling or slowly rotating approximation is violated. In this sense, we have only shown the non-existence of the {\emph{perturbative}} non-trivial, 2nd-rank Killing tensor in the new BH solution. However, we emphasize that, in this paper, we only focus on the situation where both of the approximations hold.

Although the geodesic equations are not exactly integrable, the new solution is sufficiently \emph{close} to the one found by Yunes and Pretorius~\cite{yunespretorius} that, except for the resonant orbits, the geodesic equations are still \textit{approximately integrable}. By the latter, we mean that when one orbit-averages, there still exists a Carter-like constant, i.e.~the ${\cal{O}}(\alpha'^{2} \chi'^{2})$ terms that spoil the existence of a Killing tensor are odd in $\omega t$, where $\omega$ is any of the fundamental frequencies of the motion, and thus, vanish upon orbit-averaging. This can be shown explicitly by applying canonical perturbation theory~\cite{goldstein} following e.g. Glampedakis and Babak~\cite{glampedakis}, as we discuss in Appendix~\ref{app:canonical}.
 
The new metric found here cannot be mapped to the new {\emph{bumpy metrics}} proposed in~\cite{vigelandnico}. This is because the latter assumed the existence of a non-trivial second-rank Killing tensor, while the solution found here does not possess it. We have tried to map the new solution to a generic deformed Lewis-Papapetrou spacetime~\cite{papapetrou} in one of the BL-type coordinates. We found that a naive mapping does not seem to work, which implies that a further coordinate transformation is probably needed. 

\subsection{Binding Energy, Kepler's Third Law, the Location of the ISCO and Curves of Zero Velocity}

From the definitions of the energy $E$ and the ($z$-component of) orbital angular momentum $L_z$, we have 
\ba
\dot{t} &=& \frac{E g_{\phi\phi}+L_z g_{t\phi}}{g_{t\phi^2}-g_{tt}g_{\phi\phi}}\,, \\
\dot{\phi} &=& - \frac{E g_{t\phi}+L_z g_{tt}}{g_{t\phi}^2-g_{tt}g_{\phi\phi}}\,, 
\ea
where the overhead dot stands for a derivative with respect to the affine parameter.
By substituting the above equations in $u^\alpha u_\alpha =-1$, with $u^\alpha$ the particle's four-velocity, we obtain
\be
g_{rr} \dot{r}^2 + g_{\theta\theta} \dot{\theta}^2 = V_{\eff}(r,\theta ;E,L_z)\,,
\ee
where the effective potential is given by
\be
V_{\eff} \equiv \frac{E^2 g_{\phi\phi} +2EL_z g_{t\phi}+L_z^2 g_{tt}}{g_{t\phi}^2-g_{tt}g_{\phi\phi}}-1\,.
\label{Veff}
\ee

For simplicity, we restrict attention to equatorial, circular orbits.
Then, $E$ and $L_z$ can be obtained from $V_{\eff}=0$ and $\partial V_{\eff}/\partial r =0$ as
\ba
\label{E}
E &=& E_{\K} + \delta E\,, \\
\label{Lz}
L_z &=& L_{z,{\K}} + \delta L_z\,.
\ea
Here, $E_{\K}$ and $L_{z,{\K}}$ are the energy and the ($z$-component of) orbital angular momentum for the Kerr background~\cite{bardeen}:
\ba
E_{\K} &\equiv & \frac{r^{3/2}-2Mr^{1/2} + a M^{1/2}}{r^{3/4}(r^{3/2}-3Mr^{1/2} + 2 a M^{1/2})^{1/2}}\,, \\
L_{z,{\K}} & \equiv & \frac{M^{1/2} (r^2 - 2 a M^{1/2} r^{1/2} + a^2)}{r^{3/4}(r^{3/2}-3Mr^{1/2} + 2 a M^{1/2})^{1/2}}\,, 
\ea
where we have defined $\phi$ to be positive in the direction of prograde orbits. This implies that negative $a$ corresponds to retrograde orbits. The CS corrections are 
\bw
\ba
\delta E &\equiv &   \frac{5}{4} \zeta \chi \frac{M^{11/2}}{r^3(r-3M)^{5/2}}  \Bigg( 1-\frac{33}{14} \frac{M}{r} - \frac{183}{140} \frac{M^2}{r^2} -\frac{603}{70} \frac{M^3}{r^3} + \frac{81}{4} \frac{M^4}{r^4} \Bigg)   \nn \\
& & -\frac{201}{7168} \zeta \chi^2 \frac{M^3}{r^{1/2}(r-3M)^{5/2}} \Bigg( 1+4 \frac{M}{r} -\frac{59315}{4221} \frac{M^2}{r^2} + \frac{38954}{4221} \frac{M^3}{r^3} + \frac{289564}{4221} \frac{M^4}{r^4} + \frac{188420}{1407} \frac{M^5}{r^5} \nn \\
& & - \frac{566500}{1407} \frac{M^6}{r^6} - \frac{27360}{67} \frac{M^7}{r^7} -\frac{61584}{67} \frac{M^8}{r^8} + \frac{96768}{67} \frac{M^9}{r^9} \Bigg) + \mathcal{O}(\alpha'^2 \chi'^3)\,,  \\
\delta L_z &\equiv &  \frac{15}{8} \zeta \chi \frac{M^5}{r^{3/2} (r-3M)^{5/2}}  \Bigg( 1-3 \frac{M}{r} - \frac{2}{5} \frac{M^2}{r^2} -6 \frac{M^3}{r^3} + \frac{108}{5} \frac{M^4}{r^4} \Bigg) \nn \\
& & -  \frac{603}{7168} \zeta \chi^2 \frac{r M^{5/2}}{(r-3M)^{5/2}}  \Bigg(1- \frac{4}{3} \frac{M}{r} -\frac{54833}{12663} \frac{M^2}{r^2} + \frac{110798}{12663} \frac{M^3}{r^3} + \frac{15100}{12663} \frac{M^4}{r^4} + \frac{369428}{4221} \frac{M^5}{r^5} \nn \\
& & - \frac{74092}{603} \frac{M^6}{r^6} + \frac{32768}{469} \frac{M^7}{r^7} - \frac{40688}{67} \frac{M^8}{r^8} + \frac{32256}{67} \frac{M^9}{r^9} \Bigg)+ \mathcal{O}(\alpha'^2 \chi'^3)\,.
\ea
\ew
When we expand $E$ and $L_z$ in powers of $M/r$, the leading-order correction to the binding energy $E_{b} \equiv E-1$ and $L_z$ are
\ba
\label{binding}
E_{b} &=& E_{b,{\K}} \left( 1+ \frac{201}{3584} \zeta \chi^2 \frac{M^2}{r^2} \right) + \mathcal{O}\left(\alpha'^2 \frac{M^4}{r^4} \right)\,, \nn \\
\\
L_z &=& L_{z,{\K}} \left( 1 - \frac{603}{7168} \zeta \chi^2 \frac{M^2}{r^2} \right) + \mathcal{O}\left(\alpha'^2 \frac{M^5}{r^5} \right)\,.
\ea
Relative to the leading-order Kerr (or Kepler) terms, the corrections are proportional to $(M/r)^2$ which are of 2PN orders. 
As before, the corrections in $E$ and $L_z$ would change if one used a different renormalization of the mass and spin, such as $\tilde{M}$ and $\tilde{a}$, but these quantities are not observable at spatial infinity. 

We can also derive the correction to Kepler's Third Law by calculating the orbital angular frequency of a test particle $\omega \equiv L_z/r^2$ to find
\be
\omega^2 = \omega^2_{\K} \left( 1 - \frac{603}{3584} \zeta \chi^2 \frac{M^2}{r^2} \right) + \mathcal{O}\left(\alpha'^2 \frac{M^6}{r^6} \right)\,,
\label{kepler}
\ee
where $\omega^2_{\K} \equiv M (r^{3/2} + a M^{1/2})^{-2}$~\cite{bardeen}.

However, the expressions for $E$, $L_z$ and $\omega$ above are not gauge invariant. The gauge invariant relation between $E$ and $\omega$ can be obtained by expanding Eqs.~\eqref{binding} and~\eqref{kepler} to 2PN order and eliminating $M/r$. The final result is 
\ba
\omega (E) &=& \frac{2 \sqrt{2}}{M} |E_b|^{3/2} \left[ 1 + \frac{9}{4} |E_b| - 8\sqrt{2} \chi |E_b|^{3/2} \right. \nn \\
& & \left. +\frac{891}{32} \left( 1+\frac{64}{297} \chi^2 -\frac{67}{2772} \zeta \chi^2 \right) |E_b|^{2}  \right]\nn \\
& & + \mathcal{O} \left[ |E_b|^4 \right] + {\cal{O}}(\alpha'^2 \chi'^{3})\,. 
\ea
and its inverse is
\ba
E (\omega) &=& 1-\frac{1}{2} (M \omega)^{2/3} + \frac{3}{8} (M \omega)^{4/3} - \frac{4 \chi}{3} (M \omega)^{5/3} \nn \\
& & + \frac{27}{16} \left( 1+\frac{8}{27} \chi^2 - \frac{67}{2016} \zeta \chi^2 \right) (M \omega)^{2} \nn \\
& & +\mathcal{O} \left[ (M \omega)^{7/3} \right] + {\cal{O}}(\alpha'^2 \chi'^{3})\,, 
\ea
To $\mathcal{O}(\alpha'^0 \chi'^0)$, this agrees with the standard PN $E$--$\omega$ relation shown in~\cite{blanchet-review}.
 
Let us now derive the correction to the location of ISCO. Substituting Eqs.~\eqref{E} and~\eqref{Lz} in Eq.~\eqref{Veff}, and then solving the equation $\partial^2 V_{\eff}/\partial r^2 =0$ for $r$, we obtain
\ba
r_{\ISCO} &=& r_{\ISCO,{\K}} + \frac{77\sqrt{6}}{5184} \zeta M \chi - \frac{9497219}{219469824} \zeta M \chi^2 \nn \\
& & + \mathcal{O}(\alpha'^2 \chi'^3)\,,
\ea
where the Kerr ISCO radius is given by~\cite{bardeen}
\be
r_{\ISCO,{\K}} \equiv M \left\{ 3+Z_2 - [ (3-Z_1) (3+Z_1+2Z_2) ]^{1/2} \right\}
\ee
with
\ba
Z_1 &\equiv& 1+ (1-\chi^2 )^{1/3} [ (1+\chi )^{1/3} + (1-\chi )^{1/3} ]\,, \nn \\
\\
Z_2 &\equiv& (3\chi^2+Z_1^2)^{1/2}\,.
\ea
The CS correction at linear order in $\chi$ agrees with that found in~\cite{yunespretorius}, while the $\mathcal{O}(\alpha'^{2} \chi'^2)$ term is new. The radial location of the ISCO, however, is not gauge invariant. A gauge invariant quantity can be obtained by calculating the angular orbital frequency $\omega_{\ISCO}$ at ISCO, which is
\begin{align}
\omega_{\ISCO} &= \omega_{\ISCO,{\K}} - \frac{77}{124416} \zeta \frac{\chi}{M} 
\\
&- \frac{2333803 \sqrt{6}}{31603654656} \zeta \frac{\chi^{2}}{M}+ {\cal{O}}(\alpha'^2 \chi'^{3})\,,
\end{align}
where $\omega_{\ISCO} = M^{1/2} (r_{\ISCO,{\K}}^{3/2} + \chi M^{3/2})^{-1}$.

For completeness, let us also compute the correction to the radiative efficiency $\eta$, which is defined by
\be
\eta \equiv 1-E(r_{\ISCO})\,.
\ee
This quantity corresponds to the maximum fraction of energy being radiated when a test particle accretes into a central BH. For Schwarzschild and extremal Kerr BHs, $\eta \sim 0.06$ and $\eta \sim 0.42$, respectively. The radiative efficiency for the new solution is then
\be
\eta = \eta_{\K}+ \frac{3673\sqrt{3}}{31352832} \zeta \chi  -\frac{8087\sqrt{2}}{48771072} \zeta \chi^2 + \mathcal{O}(\alpha'^2 \chi'^3)\,.
\ee
Notice that there are both linear in $\chi$ and quadratic in $\chi$ corrections. 

\begin{figure*}[htb]
\begin{center}
\begin{tabular}{c}
\includegraphics[width=5cm,clip=true]{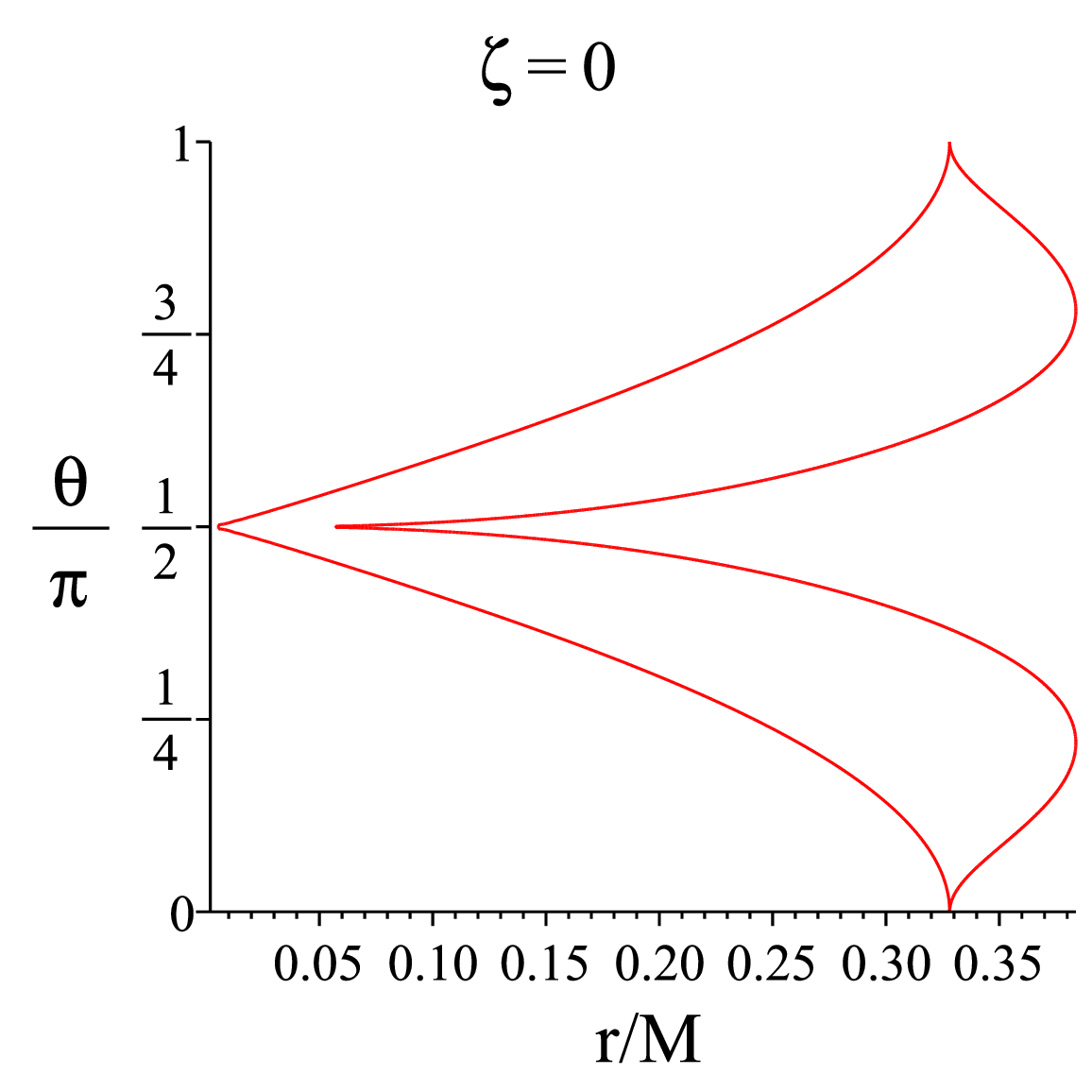}  
\hspace{1.5cm}
\includegraphics[width=5cm,clip=true]{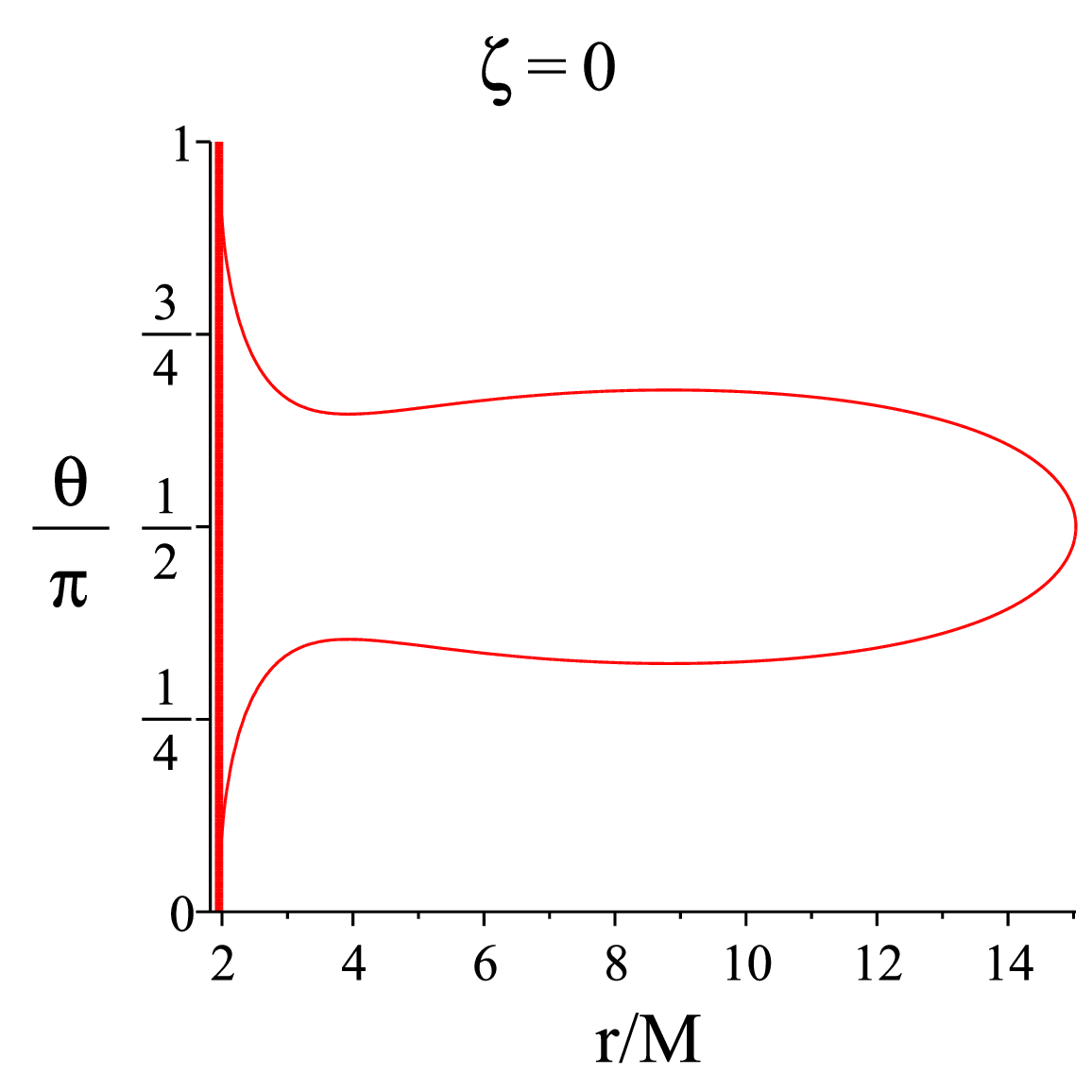} 
\end{tabular}
\caption{\label{fig:CZV} Curves of zero velocity, $V_{\eff}=0$ for the Kerr metric. The enclosed regions show the allowed orbit region $V_{\eff} \geq 0$ for $E=0.95, L_z=3M$ and $\chi =0.3$. The left panel corresponds to the inner region while the right panel corresponds to the outer region. The thick solid lines at $r/M=1.955$ correspond to the location of the event horizon when $\chi = 0.3$.}
\end{center}
\end{figure*}

\begin{figure*}[htb]
\begin{center}
\includegraphics[width=5cm,clip=true]{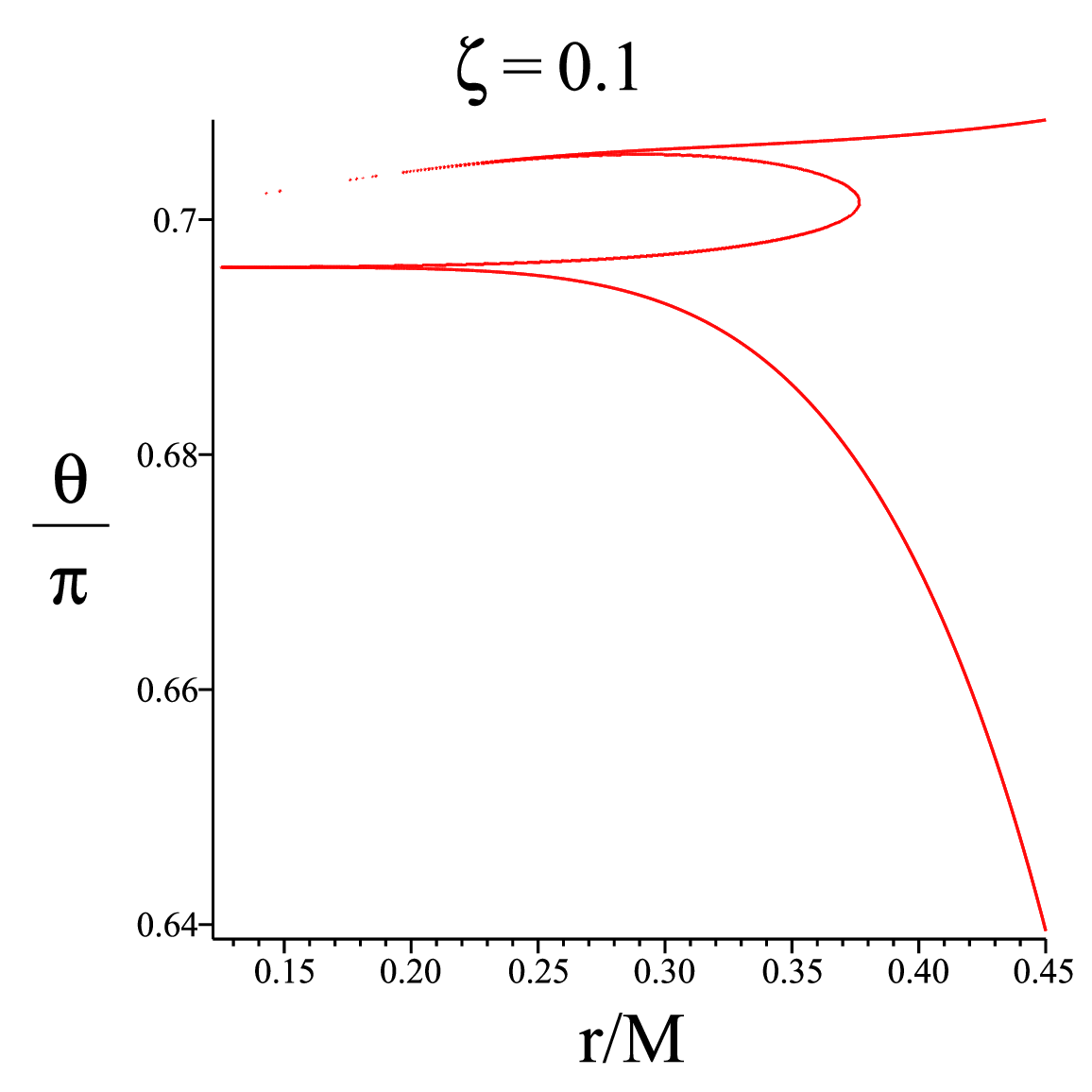} 
\hspace{1.0cm}
\includegraphics[width=5cm,clip=true]{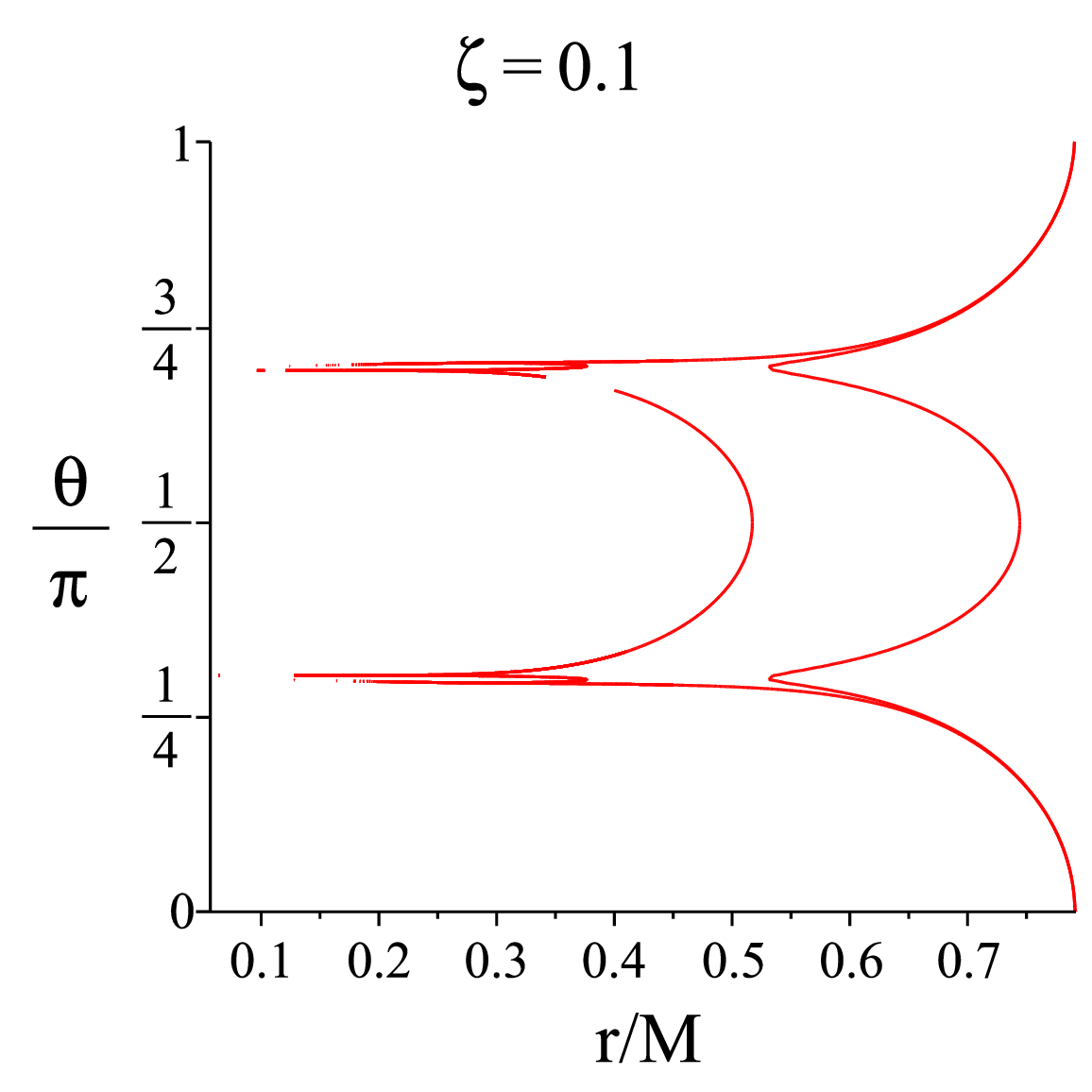}  
\hspace{1.0cm}
\includegraphics[width=5cm,clip=true]{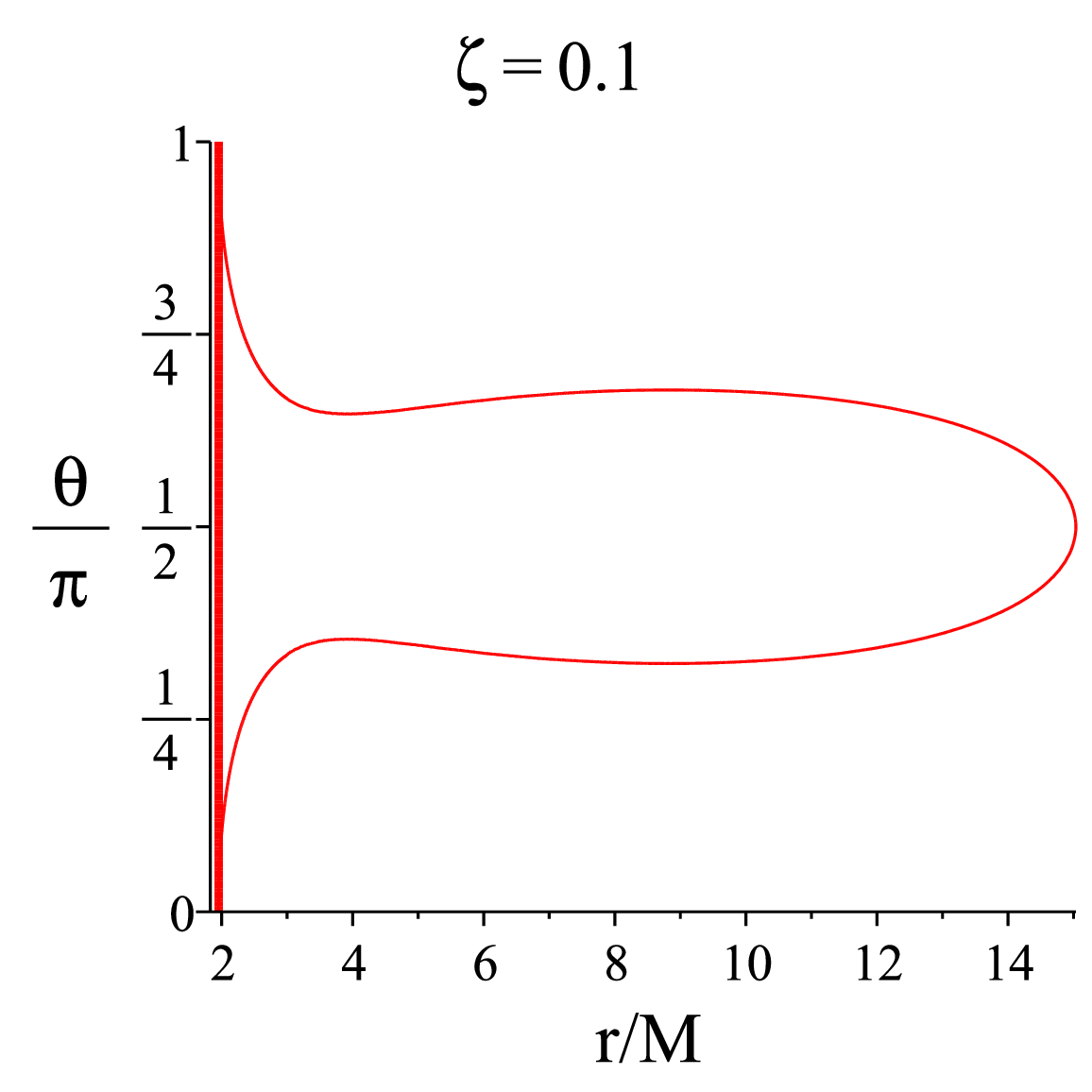}
\caption{\label{fig:CZV2} Curves of zero velocity, $V_{\eff}=0$ for the new BH metric with $\zeta=0.1$. As in Fig.~\ref{fig:CZV}, the enclosed regions show the allowed orbit region $V_{\eff} \geq 0$ for the same parameters. The middle panel corresponds to the inner region while the right panel corresponds to the outer region. The left panel zooms into the region around $\theta \approx (3/4)\pi$. Similar structure appears at around $\theta \approx (1/4)\pi$. The shaded areas are the allowed regions where test-particle orbits exist. The thick solid lines at $r/M=1.955$ correspond to the location of the event horizon when $\chi = 0.3$.}
\end{center}
\end{figure*}

Finally, let us consider curves of zero velocity (CZV)~\cite{gairbumpy,bambibarausse-CZV} in the $r-\theta$ plane, i.e. curves with $V_{\eff}=0$. Since the left-hand side of Eq.~\eqref{Veff} is always positive, bound orbits are allowed if and only if $V_{\eff} \geq 0$. Figure~\ref{fig:CZV} shows CZVs for the Kerr solution while Fig.~\ref{fig:CZV2} shows the one for the new solution. The enclosed regions represent the regions where $V_{\eff} \geq 0$ and the solid lines at $r/M=1.95$ correspond to the location of the event horizon for the particular case considered in the figures, i.e.~$E=0.95, L_z=3M, \chi =0.3$ and in the CS case, $\zeta=0.1$. When we draw these figures, we first expand the metric $g_{\mu\nu}$ in $a$ and calculate $V_{\eff}$. We do not further expand this $V_{\eff}$ in $a$, since if we do this, $V_{\eff}$ is proportional to a negative power of $a$, which would render the expansion invalid near the horizon.

Two allowed regions are clearly visible in these figures: an outer region and an inner region, for both the GR and CS cases. The outer regions are similar in GR and CS theory, although due to the scale of the figure the differences look small. As shown in~\cite{sopuertayunes} and recently in~\cite{prisgair}, orbits in the outer region are still distinguishable with gravitational wave observations. On the other hand, the structure of the inner regions change drastically, as expected since the CS correction modifies the strong field regime. These inner enclosed regions are inside the horizon, however, and thus they cannot be probed with gravitational waves, at least for slowly rotating BHs. Also, we cannot trust the perturbative solution there.

\section{Conclusions and discussions}
\label{sec:Discussions}

We have found a stationary, axisymmetric BH solution in the small-coupling and slow-rotation approximations at linear order in the coupling constant but at next-to-leading order in the spin. This solution does not satisfy the vacuum Einstein equations but the modified field equations. We used a novel technique to find this solution, based on Schwarzschild BH perturbation theory. That is, we decomposed the metric perturbation and the source terms (that come from modifications to GR) in tensor spherical harmonics, reducing the field equations to a set of coupled, ordinary differential equations that are much simpler to solve. We found that corrections at quadratic order in the spin appear in the even parity sector of the metric.

The method presented here could be used to find solutions both to higher order in $\chi$ (or $a/M)$ and to higher order in $\zeta$ defined in Eq.~\eqref{zeta}. The dynamical CS action, however, is a linear-order-in-$\zeta$ truncation of a more fundamental theory, and thus, it is valid only to linear order in the coupling constant. If one were to carry out this calculation to $\mathcal{O}(\alpha'^4 \chi')$, one would find a modification only in the gravitomagnetic sector, which is easy to compute.

A nontrivial property of the new solution is that, although it is of Petrov type D to $\mathcal{O}(\alpha'^2 \chi')$, it is of Petrov type I to $\mathcal{O}(\alpha'^2 \chi'^2)$. This is different from the Kerr metric, which is of Petrov type D to all orders in $\chi'$. The new metric does not possess a second-order Killing tensor or a Carter-like constant. This then implies that there does not exist a spacetime coordinate transformation that leads to Hamilton-Jacobi equations being separable, which also implies that geodesic motion is, in all likelihood, chaotic when corrections of $\mathcal{O}(\alpha'^2 \chi'^2)$ are included in the metric. However, although there is no exact Carter-like constant, we have also showed that the geodesic equations are still separable after orbit-averaging, except for the resonant orbits, by applying canonical perturbation theory~\cite{goldstein,glampedakis,vigelandhughes}. In some sense, then, it might be possible to recover geodesic regularity on average, although it is not clear what this means precisely. Possible future work could concentrate on studying whether geodesics in this background are truly chaotic, and if so, whether such chaos manifests itself outside the event horizon. Moreover, one could also investigate how large the shifts in the orbital frequencies of geodesic test-particles are and discuss observational prospects of probing such a spacetime (see Vigeland and Hughes~\cite{vigelandhughes} for similar work on bumpy spacetimes, as well as Appendix~\ref{app:canonical}.).

Some insight might be gained by comparing geodesic orbits in this new background to those in the Manko-Novikov (MN) spacetime~\cite{mn}. Gair \textit{et al.}~\cite{gairbumpy} investigated geodesic orbits in the MN background through CZVs in cylindrical coordinates $(t,\rho,z,\phi)$. They found that when the quadrupole moment deviates from Kerr, chaotic islands arise in the inner region of the $\rho -z$ plane. Strictly speaking, the Hamilton-Jacobi equations are not separable in the MN metric, but in the outer region, it seems that there exists a nearly invariant quantity that corresponds to a Carter-like constant. Such a result is related to the Kolmogorov, Arnold and Moser (KAM) theorem~\cite{goldstein} which states that when a separable Hamiltonian system is weakly perturbed, the perturbed motion within the phase space remains mostly in the neighborhood of the invariant tori (see related work by Apostolatos \textit{et al.}~\cite{apostolatos,gerakopoulos,contopoulos}). This suggests that except for certain resonant orbits, there should exists a fourth constant of motion in the phase space where motion is non-chaotic. This constant of motion, however, is not related to the symmetries of the spacetime anymore, and since it does not exist for resonant orbits, one cannot expect to find a global Killing tensor. In Fig.~\ref{fig:CZV}, we have shown that there are no additional CZV islands produced in the new spacetime compared to Kerr outside the event horizon. Although the structure of the inner region changes drastically, this occurs inside the horizon. Whether chaotic orbit exists in the new spacetime outside the horizon needs to be verified numerically.

In addition to the global structure of the spacetime, we investigated other properties in detail. We first found that the horizon radius and the location of the ergosphere are modified from the GR expectation, although they are not modified up to linear-order in the spin. We also computed the 2PN conservative corrections to the binding energy and Kepler's Third Law, in terms of the asymptotic mass $M$ and spin $a$. This, in addition to the dissipative corrections found in~\cite{quadratic}, will allow for the consistent calculation of the gravitational waveform to 2PN order, after the dipole-dipole scalar force is calculated. This waveform can then be mapped to the parameterized-post-Einsteinian framework~\cite{PPE} and future gravitational wave constraints on $\zeta$ can be investigated.

One might wonder whether constraints on $\zeta$ can be placed with observations of the orbital decay of binary pulsars. The gravitational fields outside of a BH, however, is very different from that outside a neutron star in dynamical CS gravity~\cite{Yunes:2009ch,alihaimoud-chen}, due to the different boundary conditions used when solving the modified field equations. Thus, the new BH solution found here cannot be used to investigate binary pulsar constraints. Instead, one would need to obtain NS solutions to quadratic order in spin, so that the dipole scalar charge that sources 2PN scalar radiation and the dipole-dipole force can be calculated. This would lead to different 2PN conservative corrections to the binding energy and Kepler Third Law. 

Two low-mass X-ray binary (LMXB) systems containing BHs have yielded a measurement or constraint on the orbital decay~\cite{kent-LMXB}, but the BH spins have not yet been determined. Since the dipole scalar charge is sourced by BH spins, it is currently impossible to place a constraint on dynamical CS gravity with these systems. Moreover, binary parameters in binary pulsars and LMXBs have been determined by assuming that GR is correct. Since the conservative part (that determines the binary parameters) appears at the same order as the dissipative part (that causes the orbital decay rate), one would have to redo the fits to simultaneously determine binary parameters and constrain $\zeta$. 

Electromagnetic radiation from accretion disks around a central BH, such as Sgr A$^*$, can also be a powerful tool to test GR~\cite{psaltis-review}. One could study how these observables, e.g.~images of BH shadows~\cite{johannsen-BHshadow,amarilla}, continuum spectrum~\cite{harko,bambibarausse-continuum}, quasi-periodic oscillations~\cite{johannsen-QPO},  Fe emission lines~\cite{johannsen-iron}, geodetic precessions and strong lensing~\cite{chen}, are modified if the central super-massive black hole (SMBH) is described by the new solution found in this paper. We expect that if the central BH is spinning moderately fast (though not extremely fast since then our slow-rotation assumption breaks down), the CS correction at quadratic order in spin will be dominant over the linear-order, CS corrections.

Another possible avenue for future work is to relax the slow-rotation approximation. Obtaining an arbitrarily fast rotating BH solution analytically seems difficult, but a numerical solution might be feasible. The results in this paper can then be used as a check of the slow-rotation limit of such a numerical solution. Once we obtain the correction to the quadrupole moment without applying the slow-rotation approximation, we can map this to a correction to the gravitational waveform.

As a final remark, one could also investigate other modified gravity theories with the techniques developed in this paper. For example, in Ref.~\cite{quadratic}, we studied Einstein-Dilaton-Gauss-Bonnet theory, where a static, spherically symmetric BH solution~\cite{yunesstein} is known, as well as a stationary, axisymmetric BH solution at linear order in spin~\cite{pani-quadratic}. We have checked that this solution can be mapped to Benenti and Francaviglia metric shown in Appendix~\ref{app:benenti}, suggesting that there exists a non-trivial second-rank Killing tensor and a Carter-like constant, and hence it admits a separability structure. One could then, for example, check whether the latter is of Petrov type D. One could also extend this solution to quadratic order in spin to see how the properties of the spacetime change compared to the linear order corrections.

\acknowledgments 
We would like to thank Leo Stein and Tsuyoshi Houri for useful comments about integrability, separability and Petrov types. We would like to also thank Dimitry Ayzenberg for pointing out a typo in the original manuscript.  
NY acknowledges support from NSF grant PHY-1114374, as well as support provided by the
National Aeronautics and Space Administration from grant NNX11AI49G, under sub-award
00001944. TT is supported by the Grant-in-Aid for Scientific Research
(Nos. 21244033, 21111006, 24111709, 24103006 and 24103001). 
Some calculations used the computer algebra-systems MAPLE, in combination with the GRTENSORII package~\cite{grtensor}.

\appendix

\section{Tensor Harmonics}
\label{app:harmonics}

In this paper, we used the following tensor spherical harmonics to decompose the metric perturbation and the source term~\cite{zerilli,sago}:
\be
\bm{a}_{\ell 0}^{(0)} = \begin{pmatrix}
Y_{\ell 0} & 0 & 0 & 0 \\
0 & 0 & 0 & 0 \\
0 & 0 & 0 & 0 \\
0 & 0 & 0 & 0 \\
\end{pmatrix}\,, 
\ee
\be
\bm{a}_{\ell 0} = \begin{pmatrix}
0 & 0 & 0 & 0 \\
0 & Y_{\ell 0}  & 0 & 0 \\
0 & 0 & 0 & 0 \\
0 & 0 & 0 & 0 \\
\end{pmatrix}\,,
\ee
\be
\bm{b}_{\ell 0} = \frac{r}{\sqrt{2\ell (\ell +1)}} \begin{pmatrix}
0 & 0 & 0 & 0 \\
0 & 0  & \frac{\partial}{\partial \theta} Y_{\ell 0} & 0 \\
0 & \frac{\partial}{\partial \theta} Y_{\ell 0} & 0 & 0 \\
0 & 0 & 0 & 0 \\
\end{pmatrix}\,,
\ee
%
%
\be
\bm{g}_{\ell 0} = \frac{r^2}{\sqrt{2}} \begin{pmatrix}
0 & 0 & 0 & 0 \\
0 & 0  & 0 & 0 \\
0 & 0 & Y_{\ell 0} & 0 \\
0 & 0 & 0 & \sin^2\theta Y_{\ell 0} \\
\end{pmatrix}\,,
\ee
\be
\bm{f}_{\ell 0} = \frac{r^2}{\sqrt{2l(\ell +1)(\ell -1)(\ell +2)}} \begin{pmatrix}
0 & 0 & 0 & 0 \\
0 & 0  & 0 & 0 \\
0 & 0 & W_{\ell 0} & 0 \\
0 & 0 & 0 & -\sin^2\theta W_{\ell 0} \\ 
\end{pmatrix}\,,
\ee
where $Y_{\ell 0}$ are the $m=0$ spherical harmonics and $W_{\ell 0}$ are given by
\be
W_{\ell 0} \equiv \left( \frac{d^2}{d\theta^2} - \cot \theta \frac{d}{d\theta} \right) Y_{\ell 0}\,.
\ee

On the other hand, the coefficients of the source after a tensor spherical harmonics decomposition are
\bw
\ba
A_{00}^{(0)} &= & \frac{25 \sqrt{\pi}}{64} \zeta \frac{M^4}{r^6} f(r) \chi^2 \left(1 +4\frac{M}{r} + \frac{322}{3} \frac{M^2}{r^2} + \frac{396}{5} \frac{M^3}{r^3} +  \frac{2092}{25} \frac{M^4}{r^4} -\frac{46656}{25} \frac{M^5}{r^5} \right)\,,  
\\
A_{20}^{(0)} &= & \frac{5 \sqrt{5\pi}}{64} \zeta \frac{M^4}{r^6} f(r) \chi^2 \left(1 +4\frac{M}{r} + \frac{1324}{15} \frac{M^2}{r^2} + \frac{912}{5} \frac{M^3}{r^3} +  308 \frac{M^4}{r^4} -\frac{48384}{25} \frac{M^5}{r^5} \right)\,,  
\\
A_{00} &= & \frac{25 \sqrt{\pi}}{192} \zeta \frac{M^4}{r^6} \frac{1}{f(r)^2} \chi^2 \left(1 +2\frac{M}{r} -54 \frac{M^2}{r^2} -32 \frac{M^3}{r^3} -\frac{1044}{175} \frac{M^4}{r^4} +\frac{18312}{25} \frac{M^5}{r^5} -\frac{24192}{25} \frac{M^6}{r^6} \right)\,,  \\
A_{20} &= & \frac{25 \sqrt{5\pi}}{192} \zeta \frac{M^4}{r^6} \frac{1}{f(r)^2} \chi^2 \left(1 +2\frac{M}{r} -\frac{492}{25} \frac{M^2}{r^2} +\frac{904}{25} \frac{M^3}{r^3} +\frac{35676}{875} \frac{M^4}{r^4} +\frac{43896}{125} \frac{M^5}{r^5} -\frac{13824}{25} \frac{M^6}{r^6} \right)\,,  
\\
B_{20} &= & -\frac{5 \sqrt{15\pi}}{48} \zeta \frac{M^4}{r^6} \frac{1}{f(r)^2} \chi^2 \left(1 + \frac{M}{r} -28 \frac{M^2}{r^2} +\frac{338}{5} \frac{M^3}{r^3} +\frac{2496}{175} \frac{M^4}{r^4} +\frac{64296}{175} \frac{M^5}{r^5} -\frac{22896}{25} \frac{M^6}{r^6} \right)\,,  
\\
G_{00}^{(s)} &= & -\frac{25 \sqrt{2\pi}}{96} \zeta \frac{M^4}{r^6} \frac{1}{f(r)^2} \chi^2 \left(1 -87 \frac{M^2}{r^2} +18 \frac{M^3}{r^3} +\frac{33892}{175} \frac{M^4}{r^4} +\frac{302136}{175} \frac{M^5}{r^5} -\frac{100176}{25} \frac{M^6}{r^6} +\frac{44928}{25} \frac{M^7}{r^7} \right)\,,  \nn \\
\\
G_{20}^{(s)} &= & -\frac{5 \sqrt{10\pi}}{48} \zeta \frac{M^4}{r^6} \frac{1}{f(r)^2} \chi^2 \left(1-\frac{219}{5}\frac{M^2}{r^2} +162 \frac{M^3}{r^3} -\frac{19868}{175} \frac{M^4}{r^4} +\frac{186072}{175} \frac{M^5}{r^5} -\frac{121416}{25} \frac{M^6}{r^6} +\frac{127872}{25} \frac{M^7}{r^7} \right)\,,  \nn \\
\\
F_{20} &= & \frac{5 \sqrt{15\pi}}{96} \zeta \frac{M^4}{r^6} \frac{1}{f(r)} \chi^2 \left(1 +2 \frac{M}{r} -\frac{272}{5} \frac{M^2}{r^2} +\frac{984}{5} \frac{M^3}{r^3} +\frac{7788}{175} \frac{M^4}{r^4} +\frac{18264}{25} \frac{M^5}{r^5} -\frac{55296}{25} \frac{M^6}{r^6} \right)\,. 
\ea
%
By substituting the above source terms in Eqs.~\eqref{Sago1}-~\eqref{Sago5}, we obtain a set of ordinary differential equations for $(H_{000},H_{200},K_{00},H_{020},H_{220},K_{20})$, which we solved to find
%
\begin{align}
H_{000} &= - \frac{5\sqrt{\pi}}{192} \zeta \chi^2 \frac{M^5}{r^5 f(r)} \left( 1 +100 \frac{M}{r} +194 \frac{M^2}{r^2} + \frac{2220}{7} \frac{M^3}{r^3} - \frac{1512}{5} \frac{M^4}{r^4} \right) + {\cal{O}}(\alpha'^2 \chi'^{3})\,, \\
H_{200} &= - \frac{25\sqrt{\pi}}{192} \zeta \chi^2 \frac{M^4}{r^4 f(r)} \left( 1 +3 \frac{M}{r} + \frac{322}{5}\frac{M^2}{r^2} + \frac{198}{5} \frac{M^3}{r^3} + \frac{6276}{175} \frac{M^4}{r^4} - \frac{17496}{25} \frac{M^5}{r^5} \right)+ {\cal{O}}(\alpha'^2 \chi'^{3})\,, \\ 
K_{00} &= {\cal{O}}(\alpha'^2 \chi'^{3})\,, \\
H_{020} &= \frac{469\sqrt{5\pi}}{5230} \zeta \chi^2 \frac{M^3}{r^3 f(r)} \left( 1 + \frac{M}{r} +\frac{4474}{4221} \frac{M^2}{r^2} - \frac{2060}{469} \frac{M^3}{r^3} + \frac{1500}{469} \frac{M^4}{r^4} - \frac{2140}{201} \frac{M^5}{r^5} + \frac{9256}{201} \frac{M^6}{r^6} - \frac{5376}{67} \frac{M^7}{r^7} \right)+ {\cal{O}}(\alpha'^2 \chi'^{3})\,, \nn \\
\\
H_{220} &= \frac{201\sqrt{5\pi}}{2240} \zeta \chi^2 \frac{M^3}{r^3} \left( 1 + \frac{1459}{603}\frac{M}{r} +\frac{20000}{4221} \frac{M^2}{r^2} + \frac{51580}{1407} \frac{M^3}{r^3} - \frac{7580}{201} \frac{M^4}{r^4} - \frac{22492}{201} \frac{M^5}{r^5} - \frac{40320}{67} \frac{M^6}{r^6} \right)+ {\cal{O}}(\alpha'^2 \chi'^{3})\,, \\
K_{20} &= \frac{201\sqrt{5\pi}}{2240} \zeta \chi^2 \frac{M^3}{r^3} \left( 1 + \frac{1420}{603}\frac{M}{r} +\frac{18908}{4221} \frac{M^2}{r^2} + \frac{1480}{603} \frac{M^3}{r^3} + \frac{22460}{1407} \frac{M^4}{r^4} + \frac{3848}{201} \frac{M^5}{r^5} + \frac{5376}{67} \frac{M^6}{r^6} \right)+ {\cal{O}}(\alpha'^2 \chi'^{3})\,.
\end{align}
\ew
This solutions can then be used to reconstruct the metric perturbation, presented in the main text.

\section{Alternative Ways to Prove the Non-admittance of a Separability Structure}
\label{app:Killing}

In Sec.~\ref{sec:separability}, we have shown that a 2nd-rank Killing tensor does not exist for the new BH solution by directly attempting to solve the Killing equations with a general ansatz for the Killing tensor, restricted by the symmetries of the spacetime. This implies that the new BH solution does not admit a separability structure. In this appendix, we verify this point in two additional ways.

\subsection{Levi-Civita test}

Whether a Hamiltonian system is separable or not in a given coordinate system $x^\alpha$ can be determined by the so-called {\emph{Levi-Civita test}}~\cite{levi-civita,yasui}. This test states that the Hamilton-Jacobi equations are separable if and only if the Hamiltonian $H \equiv (1/2) g^{\mu\nu} p_\mu p_\nu$~\cite{carter} satisfies the following equations for \textit{each pair} of \textit{distinct} indices $\alpha$ and $\beta$:
\ba
& & \frac{\partial H}{\partial x^\alpha} \frac{\partial H}{\partial x^\beta}\frac{\partial^2 H}{\partial p_\alpha p_\beta} +\frac{\partial H}{\partial p_\alpha} \frac{\partial H}{\partial p_\beta}\frac{\partial^2 H}{\partial x^\alpha x^\beta} \nn \\
& & - \frac{\partial H}{\partial p_\alpha} \frac{\partial H}{\partial x^\beta}\frac{\partial^2 H}{\partial x^\alpha p_\beta} - \frac{\partial H}{\partial x^\alpha} \frac{\partial H}{\partial p_\beta}\frac{\partial^2 H}{\partial p_\alpha x^\beta} = 0\,.
\label{LC-test}
\ea
Here, $p_\alpha$ is the conjugate momentum of the coordinate $x^\alpha$. Note that the repeated indices do \textit{not} indicate summation. For the stationary and axisymmetric spacetime, since the metric only depends on $r$ and $\theta$, the only relevant combination of $(\alpha,\beta)$ in the above equation is $(\alpha,\beta)=(r, \theta)$. We have checked that the slowly rotating BH solution in dynamical CS gravity at linear order in spin satisfies Eq.~\eqref{LC-test}. The new solution in BL-like coordinates does not satisfy the above equations, as we verified explicitly with symbolic manipulation software. This suggests that there is no conserved quantity corresponding to the Carter constant that can be constructed from a 2nd-order Killing tensor, and thus, the Hamilton-Jacobi equations are not separable in BL-like coordinates. 

Next, we investigated whether there exist any coordinates in which the new BH solution satisfies Eq.~\eqref{LC-test}. In particular, we allowed for diffeomorphisms $x^{\mu} \rightarrow x'^{\mu} = x^{\mu} + \xi^{\mu}$, restricted to $\xi^{\mu}$ being of order $\mathcal{O}(\alpha'^2 \chi'^2)$. With this coordinate transformation, $g_{\mu\nu}$ transforms via $g_{\mu\nu} \rightarrow g'_{\mu\nu} = g_{\mu\nu} - 2 h_{(\mu ;\nu)}$. We verified, however, that the transformed Hamiltonian does not satisfy Eq.~\eqref{LC-test} with $(\alpha,\beta)=(r,\theta)$. We found that the left-hand-side of this equation contains only five terms, which are proportional to the combinations $(E^2 p_r p_\theta)$, $(E L_z p_r p_\theta)$, $(L_z^2 p_r p_\theta)$, $(p_r^3 p_\theta)$ and $(p_r p_\theta^3)$. If Eq.~\eqref{LC-test} is satisfied, the coefficient of each term must vanish separately (i.e. there are five equations to be satisfied). On the other hand, we found that $\xi^{\mu}$ appears in these coefficients only through $\partial_\theta \xi^r$, $\partial_r \xi^\theta$, $\partial_r \partial_\theta \xi^r$ and $\partial_r \partial_\theta \xi^\theta$. We tried to solve the five equations for these four quantities, but could not obtain a consistent solution, which then proves that there does not exist any coordinate transformation where Eq.~\eqref{LC-test} is satisfied.

\subsection{Mapping to the General Metric that admits a Separability Structure}
\label{app:benenti}

The inverse of the metric components of a spacetime that admits a separability structure can be expressed as~\cite{benenti}
\ba
g^{rr} & = & \frac{Q(r)}{r^2 + p^2}, \qquad g^{\theta\theta} = \frac{P(p)}{(r^2+p^2) a^2 \sin^2 \theta}\,, \nn \\
g^{ab} &=& \frac{Q(r)}{r^2 + p^2} \zeta_r^{ab}(r) + \frac{P(p)}{r^2 + p^2} \zeta_p^{ab}(p), \quad (a,b = t,\phi)\,, \nn \\
\ea
where $p \equiv a \cos\theta$. There are four functions of $r$, $Q(r)$ and $\zeta_r^{ab}(r)$, and four functions of $\theta$, $P(p)$ and $\zeta_p^{ab}(p)$. The Kerr solution can be expressed as
\ba
Q_{\K}(r) &=& \Delta, \qquad P_{\K}(p) = a^2 \sin^2\theta\,, \nn \\
\zeta_{r,{\K}}^{ab}(r)& = & \frac{1}{\Delta^2} \begin{pmatrix} (r^2+a^2)^2 & a(r^2+a^2) \\
                                                   a(r^2+a^2) & a^2  
                               \end{pmatrix}\,, \nn \\
\zeta_{p,{\K}}^{ab}(p)& = &  \begin{pmatrix} a^2 \sin^2 \theta & a \\
                                                   a & \frac{1}{\sin^2\theta}  
                               \end{pmatrix}\,.         
\ea

The deviation from the Kerr solution can be parametrized as
\ba
g^{rr} &=& \frac{Q_{\K}}{r^2+p^2} (1+\delta Q)\,, \\
g^{\theta\theta} &=& \frac{P_{\K}}{(r^2+p^2)a^2\sin^2\theta} (1+\delta P)\,, \\
g^{tt} &=& \frac{1}{r^2+p^2} \left[ Q_{\K} \zeta _{r,{\K}}^{tt} (1+ \delta Q + \delta \zeta _r^{tt}) \right.\nn \\
& & \left. +P_{\K} \zeta _{p,{\K}}^{tt}(1+ \delta P + \delta \zeta _p^{tt}) \right]\,, \\
g^{t\phi} &=& \frac{1}{r^2+p^2} \left[ Q_{\K} \zeta _{r,{\K}}^{t\phi} (1+ \delta Q + \delta \zeta _r^{t\phi}) \right.\nn \\
& & \left. +P_{\K} \zeta _{p,{\K}}^{t\phi}(1+ \delta P + \delta \zeta _p^{t\phi}) \right]\,, \\
g^{\phi\phi} &=& \frac{1}{r^2+p^2} \left[ Q_{\K} \zeta _{r,{\K}}^{\phi\phi} (1+ \delta Q + \delta \zeta _r^{\phi\phi}) \right.\nn \\
& & \left. +P_{\K} \zeta _{p,{\K}}^{\phi\phi}(1+ \delta P + \delta \zeta _p^{\phi\phi}) \right]\,. 
\ea
where, 
\be
\delta A \equiv \frac{A-A_{\K}}{A_{\K}}\,, \qquad (A = Q,P,\zeta_r^{ab},\zeta_p^{ab})\,.
\ee
This can be interpreted as the most general bumpy spacetime that admits a separability structure.

Let us now try to map the new BH solution to the above bumpy metric. First, to $\mathcal{O}(\alpha'^2 \chi')$, the only relevant parameters are $\delta Q$, $\delta P$, $\delta \zeta_r^{tt}$ and $\delta \zeta_r^{t\phi}$. We find that 
\ba
\delta Q &=& \delta P = \delta \zeta_r^{tt} = 0\,, \nn \\
\delta \zeta_r^{t\phi} &=& -\frac{5}{8} \zeta \frac{M^4}{r^4} \left( 1+\frac{12}{7} \frac{M}{r} + \frac{27}{10} \frac{M^2}{r^2}  \right)\,.
\ea
To $\mathcal{O}(\zeta \chi^2)$, we find that the new BH solution cannot be mapped to the above bumpy spacetime with BL-like coordinates. We also considered coordinate transforming the new BH solution to find a map to the above metric. Once again, however, we found that there does not exist any coordinates where the new BH solution can be mapped to the above bumpy spacetime. This then implies that new BH solution does not admit a separability structure.  

\section{Canonical perturbation theory and secular separability of the geodesic equations}
\label{app:canonical}

Consider a Hamiltonian system in angle-action variables, where we use $w^{\alpha}$ to denote the angle variable and $I^\alpha = (p_t,J_r,J_\theta,J_\phi)$ to denote the conjugate action variables, with 
\be
J_i \equiv \oint dx^i p_i\,.
\ee
We can then express the Hamiltonian as
\ba
H &=& \frac{1}{2} g^{\mu\nu}_{\YP} p_\mu p_\nu +\frac{1}{2}  \epsilon  h^{\mu\nu} p_\mu p_\nu + \mathcal{O}(\epsilon^2) \nn \\
& =& H_0(I_\beta ) + \epsilon H_1(w^\alpha, I_\beta )+ \mathcal{O}(\epsilon^2)\,, 
\ea
where $g^{\mu\nu}_\mrm{YP}$ denotes the stationary, axisymmetric CS BH metric at $\mathcal{O}(\alpha'^2 \chi')$ found by Yunes and Pretorius~\cite{yunespretorius} and $h^{\mu\nu}$ is the difference between the new solution and theirs. We have also defined $H_0 \equiv (1/2) g^{\mu\nu}_{\YP} p_\mu p_\nu$ and $H_1 \equiv (1/2) h^{\mu\nu} p_\mu p_\nu$, where $\epsilon = \mathcal{O}(\chi')$ is a bookkeeping parameter that labels the order of the perturbation. Clearly, the first (second) term is the unperturbed (perturbed) Hamiltonian. Note that the unperturbed Hamiltonian depends only on the action variables, while the perturbed one depends both on the angle and action variables.

Now, let us seek a canonical transformation from $(w^\alpha, I_\beta)$ to $(\hat{w}^\alpha, \hat{I}_\beta)$ such that the new action $\hat{I}_\beta$ is constant and the new angle variables $\hat{w}^\alpha$ are linear in the affine parameter. With this choice, the new Hamiltonian would depend only on the new action variables. Such a canonical transformation exists if one can find an appropriate generating function, which can be parameterized as~\cite{goldstein,glampedakis}
\be
F(w^\alpha,\hat{I}_\beta) = w^\alpha \hat{I}_\alpha + \Phi (w^\alpha, \hat{I}_\beta)\,,
\ee
where $\Phi$ is a function of the old angle variables and the new action variables, which are assumed to be periodic in $w^\alpha$. 

Whether such a canonical transformation exists then depends on whether $\Phi$ exists. This function must satisfy the Hamilton-Jacobi equation
\be
H\left( w^\alpha,\frac{\partial F}{\partial w^\alpha} \right) = \hat{H} (\hat{I_\beta}, \epsilon)\,,
\label{HJ}
\ee
where $\hat{H}$ is the new Hamiltonian. We expand $\hat{H}$ as
\be
\hat{H}(\hat{I}_\beta, \epsilon) = H_0(\hat{I}_\beta) + \epsilon H_1(\hat{I}_\beta) + \mathcal{O}(\epsilon^2)\,,
\ee
and substitute this into Eq.~\eqref{HJ} to yield
\ba
H_0(\hat{I}_\beta) &=& \hat{H}_0(\hat{I}_\beta)\,, \\
\nu^\alpha_0 \frac{\partial \Phi}{\partial w^\alpha} + H_1(w^\alpha, \hat{I}_\beta) & = & \hat{H}_1(\hat{I}_\beta)\,,
\label{HJ2}
\ea
where $\nu^\alpha_0 \equiv \partial H_0/ \partial I_\alpha$ represents the unperturbed frequencies.
Next, since $\Phi$ is assumed to be periodic in $w^\alpha$, we Fourier decompose it
\be
\Phi (w^\alpha, \hat{I}_\beta) = \sum_{\bm{j}} B_{\bm{j}} (\hat{I}_\beta) e^{2 \pi i (\bm{j} \cdot \bm{w})}\,,
\label{Phi-decomp}
\ee
where $\bm{j}$ is a 3-dimensional vector of the integer indices. The coefficients $B_{\bm{j}}$ can be determined by solving Eq.~\eqref{HJ2}. Since the derivative of $\Phi$ with respect to $w^\alpha$ does not contain any constant term, the first term on the right hand side of Eq.~\eqref{HJ2} does not depend on $\hat{I}_\beta$ when one orbit-averages. Therefore, upon orbit-averaging, Eq.~\eqref{HJ2} becomes 
\be
\left \langle H_1 \right \rangle (w^\alpha,\hat{I}_\beta)  = \hat{H}_1(\hat{I}_\beta)\,,
\ee
where the orbital averaging is defined as
\be
\left \langle A \right \rangle \equiv \oint dw^\alpha A\,,
\ee
for any quantity $A$. The oscillatory part of Eq.~\eqref{HJ2} must satisfy 
\be
\nu^\alpha_0 \frac{\partial \Phi}{\partial w^\alpha} = \left \langle H_1 \right \rangle - H_1\,.
\label{HJ3}
\ee
Similar to Eq.~\eqref{Phi-decomp}, we decompose the right hand side of the above equation as
\be
\left \langle H_1 \right \rangle - H_1 =  \sum_{\bm{j \neq 0}} C_{\bm{j}} (\hat{I}_\beta) e^{2 \pi i \bm{j} \cdot \bm{w}}\,.
\ee
By substituting this equation and Eq.~\eqref{Phi-decomp} into Eq.~\eqref{HJ3}, we can determine the coefficients $B_{\bm{j}}$ in Eq.~\eqref{Phi-decomp} as
\be
B_{\bm{j}} (\hat{I}_\beta) = \frac{C_{\bm{j}} (\hat{I}_\beta)}{2 \pi i (\bm{j} \cdot \bm{\nu}_0)}, \quad \bm{j} \neq 0\,.
\label{B}
\ee
Thus, when $\bm{j} \cdot \bm{\nu}_0 \neq 0$, the $B_{\bm{j}}$ coefficients exist, which then implies that $\Phi$ and the canonical transformation in question indeed exist.

If the generating function exists, the relation between the old and the new action variables is
\be
I_\alpha = \frac{\partial F}{\partial w^\alpha} = \hat{I}_\alpha + \frac{\partial \Phi}{\partial w^\alpha}\,.
\ee
Taking the orbital average of this equation, we obtain 
\be
\left\langle I_\beta \right\rangle = \left\langle \hat{I}_\beta \right\rangle\,.
\ee
Thus, there is no secular change in the action variables, which means that the conserved quantities (energy, angular momentum and the Carter-like constant) are not modified in a secular sense~\cite{glampedakis}. Therefore, when we take an orbital average, the geodesic equations must remain \textit{secularly} integrable.

This, however, does not mean that there are no secular changes in the angular frequencies $\omega^i$.
The shifts in these frequencies conjugate to the orbital proper time can be calculated via~\cite{vigelandhughes}
\be
\delta \omega^i \equiv \frac{1}{m} \frac{\partial \left\langle H_1 \right\rangle}{\partial I_i}\,,
\ee
where $m$ is the mass of the test particle. This can be related to the shifts in the \textit{observed} angular frequencies $\Omega^i$ via
\be
\delta \Omega^i = \frac{\delta \omega^i}{\Gamma} - \frac{\omega^i \delta \Gamma}{\Gamma^2}
\ee
with 
\be
\omega^i \equiv \frac{1}{m} \frac{\partial H_0}{\partial I_i}, \quad \Gamma \equiv \frac{1}{m} \frac{\partial H_0}{\partial I_t}, \quad \delta \Gamma \equiv \frac{1}{m} \frac{\partial \left\langle H_1 \right\rangle}{\partial I_t}\,.
\ee

Resonant orbits must be treated separately. In this case, $\bm{j} \cdot \bm{\nu}_0 =0$ and the right-hand side of Eq.~\eqref{B} diverges. This, in turn, means a canonical transformation does not exist, and thus, a Carter-like constant of motion does not exist either at resonance, even when one takes the orbit averaging. Since a constant of the motion must be a global quantity, this implies that a Carter constant cannot globally exist. 

\bibliography{master}

\begin{thebibliography}{77}
\expandafter\ifx\csname natexlab\endcsname\relax\def\natexlab#1{#1}\fi
\expandafter\ifx\csname bibnamefont\endcsname\relax
  \def\bibnamefont#1{#1}\fi
\expandafter\ifx\csname bibfnamefont\endcsname\relax
  \def\bibfnamefont#1{#1}\fi
\expandafter\ifx\csname citenamefont\endcsname\relax
  \def\citenamefont#1{#1}\fi
\expandafter\ifx\csname url\endcsname\relax
  \def\url#1{\texttt{#1}}\fi
\expandafter\ifx\csname urlprefix\endcsname\relax\def\urlprefix{URL }\fi
\providecommand{\bibinfo}[2]{#2}
\providecommand{\eprint}[2][]{\url{#2}}

\bibitem[{\citenamefont{Will}(2006)}]{will-living}
\bibinfo{author}{\bibfnamefont{C.~M.} \bibnamefont{Will}},
  \bibinfo{journal}{Living Reviews in Relativity} \textbf{\bibinfo{volume}{9}}
  (\bibinfo{year}{2006}), \eprint{gr-qc/0510072},
  \urlprefix\url{http://www.livingreviews.org/lrr-2006-3}.

\bibitem[{\citenamefont{Robinson}(1975)}]{robinson}
\bibinfo{author}{\bibfnamefont{D.}~\bibnamefont{Robinson}},
  \bibinfo{journal}{Phys.Rev.Lett.} \textbf{\bibinfo{volume}{34}},
  \bibinfo{pages}{905} (\bibinfo{year}{1975}).

\bibitem[{\citenamefont{Israel}(1967)}]{israel}
\bibinfo{author}{\bibfnamefont{W.}~\bibnamefont{Israel}},
  \bibinfo{journal}{Phys. Rev.} \textbf{\bibinfo{volume}{164}},
  \bibinfo{pages}{1776} (\bibinfo{year}{1967}).

\bibitem[{\citenamefont{Israel}(1968)}]{israel2}
\bibinfo{author}{\bibfnamefont{W.}~\bibnamefont{Israel}},
  \bibinfo{journal}{Commun.Math.Phys.} \textbf{\bibinfo{volume}{8}},
  \bibinfo{pages}{245} (\bibinfo{year}{1968}).

\bibitem[{\citenamefont{Hawking}(1971)}]{hawking-uniqueness0}
\bibinfo{author}{\bibfnamefont{S.}~\bibnamefont{Hawking}},
  \bibinfo{journal}{Phys.Rev.Lett.} \textbf{\bibinfo{volume}{26}},
  \bibinfo{pages}{1344} (\bibinfo{year}{1971}).

\bibitem[{\citenamefont{Hawking}(1972)}]{hawking-uniqueness}
\bibinfo{author}{\bibfnamefont{S.~W.} \bibnamefont{Hawking}},
  \bibinfo{journal}{Commun. Math. Phys.} \textbf{\bibinfo{volume}{25}},
  \bibinfo{pages}{152} (\bibinfo{year}{1972}).

\bibitem[{\citenamefont{Carter}(1971)}]{carter-uniqueness}
\bibinfo{author}{\bibfnamefont{B.}~\bibnamefont{Carter}},
  \bibinfo{journal}{Phys.Rev.Lett.} \textbf{\bibinfo{volume}{26}},
  \bibinfo{pages}{331} (\bibinfo{year}{1971}).

\bibitem[{\citenamefont{Yunes and
  Pretorius}(2009{\natexlab{a}})}]{yunespretorius}
\bibinfo{author}{\bibfnamefont{N.}~\bibnamefont{Yunes}} \bibnamefont{and}
  \bibinfo{author}{\bibfnamefont{F.}~\bibnamefont{Pretorius}},
  \bibinfo{journal}{Phys. Rev.} \textbf{\bibinfo{volume}{D79}},
  \bibinfo{pages}{084043} (\bibinfo{year}{2009}{\natexlab{a}}),
  \eprint{0902.4669}.

\bibitem[{\citenamefont{Konno et~al.}(2009)\citenamefont{Konno, Matsuyama, and
  Tanda}}]{konnoBH}
\bibinfo{author}{\bibfnamefont{K.}~\bibnamefont{Konno}},
  \bibinfo{author}{\bibfnamefont{T.}~\bibnamefont{Matsuyama}},
  \bibnamefont{and} \bibinfo{author}{\bibfnamefont{S.}~\bibnamefont{Tanda}},
  \bibinfo{journal}{Prog.Theor.Phys.} \textbf{\bibinfo{volume}{122}},
  \bibinfo{pages}{561} (\bibinfo{year}{2009}), \eprint{0902.4767}.

\bibitem[{\citenamefont{Yunes and Stein}(2011)}]{yunesstein}
\bibinfo{author}{\bibfnamefont{N.}~\bibnamefont{Yunes}} \bibnamefont{and}
  \bibinfo{author}{\bibfnamefont{L.~C.} \bibnamefont{Stein}},
  \bibinfo{journal}{Phys. Rev.} \textbf{\bibinfo{volume}{D83}},
  \bibinfo{pages}{104002} (\bibinfo{year}{2011}), \eprint{1101.2921}.

\bibitem[{\citenamefont{Pani et~al.}(2011)\citenamefont{Pani, Macedo, Crispino,
  and Cardoso}}]{pani-quadratic}
\bibinfo{author}{\bibfnamefont{P.}~\bibnamefont{Pani}},
  \bibinfo{author}{\bibfnamefont{C.~F.~B.} \bibnamefont{Macedo}},
  \bibinfo{author}{\bibfnamefont{L.~C.~B.} \bibnamefont{Crispino}},
  \bibnamefont{and} \bibinfo{author}{\bibfnamefont{V.}~\bibnamefont{Cardoso}},
  \bibinfo{journal}{Phys. Rev.} \textbf{\bibinfo{volume}{D84}},
  \bibinfo{pages}{087501} (\bibinfo{year}{2011}), \eprint{1109.3996}.

\bibitem[{\citenamefont{Psaltis}(2008)}]{psaltis-review}
\bibinfo{author}{\bibfnamefont{D.}~\bibnamefont{Psaltis}},
  \bibinfo{journal}{Living Reviews in Relativity} \textbf{\bibinfo{volume}{11}}
  (\bibinfo{year}{2008}), \eprint{0806.1531},
  \urlprefix\url{http://www.livingreviews.org/lrr-2008-9}.

\bibitem[{\citenamefont{Yunes et~al.}(2011)\citenamefont{Yunes, Kocsis, Loeb,
  and Haiman}}]{Yunes:2011ws}
\bibinfo{author}{\bibfnamefont{N.}~\bibnamefont{Yunes}},
  \bibinfo{author}{\bibfnamefont{B.}~\bibnamefont{Kocsis}},
  \bibinfo{author}{\bibfnamefont{A.}~\bibnamefont{Loeb}}, \bibnamefont{and}
  \bibinfo{author}{\bibfnamefont{Z.}~\bibnamefont{Haiman}},
  \bibinfo{journal}{Phys.Rev.Lett.} \textbf{\bibinfo{volume}{107}},
  \bibinfo{pages}{171103} (\bibinfo{year}{2011}), \eprint{1103.4609}.

\bibitem[{\citenamefont{Kocsis et~al.}(2011)\citenamefont{Kocsis, Yunes, and
  Loeb}}]{Kocsis:2011dr}
\bibinfo{author}{\bibfnamefont{B.}~\bibnamefont{Kocsis}},
  \bibinfo{author}{\bibfnamefont{N.}~\bibnamefont{Yunes}}, \bibnamefont{and}
  \bibinfo{author}{\bibfnamefont{A.}~\bibnamefont{Loeb}},
  \bibinfo{journal}{Phys.Rev.} \textbf{\bibinfo{volume}{D84}},
  \bibinfo{pages}{024032} (\bibinfo{year}{2011}), \eprint{1104.2322}.

\bibitem[{\citenamefont{Ryan}(1995)}]{Ryan:1995wh}
\bibinfo{author}{\bibfnamefont{F.~D.} \bibnamefont{Ryan}},
  \bibinfo{journal}{Phys. Rev.} \textbf{\bibinfo{volume}{D52}},
  \bibinfo{pages}{5707} (\bibinfo{year}{1995}).

\bibitem[{\citenamefont{Ryan}(1997)}]{Ryan:1997hg}
\bibinfo{author}{\bibfnamefont{F.~D.} \bibnamefont{Ryan}},
  \bibinfo{journal}{Phys. Rev.} \textbf{\bibinfo{volume}{D56}},
  \bibinfo{pages}{1845} (\bibinfo{year}{1997}).

\bibitem[{\citenamefont{Barack and Cutler}(2007)}]{Barack:2006pq}
\bibinfo{author}{\bibfnamefont{L.}~\bibnamefont{Barack}} \bibnamefont{and}
  \bibinfo{author}{\bibfnamefont{C.}~\bibnamefont{Cutler}},
  \bibinfo{journal}{Phys. Rev.} \textbf{\bibinfo{volume}{D75}},
  \bibinfo{pages}{042003} (\bibinfo{year}{2007}), \eprint{gr-qc/0612029}.

\bibitem[{\citenamefont{Glampedakis}(2005)}]{glampedakis-emri}
\bibinfo{author}{\bibfnamefont{K.}~\bibnamefont{Glampedakis}},
  \bibinfo{journal}{Class. Quant. Grav.} \textbf{\bibinfo{volume}{22}},
  \bibinfo{pages}{S605} (\bibinfo{year}{2005}), \eprint{gr-qc/0509024}.

\bibitem[{\citenamefont{Amaro-Seoane et~al.}(2007)}]{emri-review}
\bibinfo{author}{\bibfnamefont{P.}~\bibnamefont{Amaro-Seoane}}
  \bibnamefont{et~al.}, \bibinfo{journal}{Class. Quant. Grav.}
  \textbf{\bibinfo{volume}{24}}, \bibinfo{pages}{R113} (\bibinfo{year}{2007}),
  \eprint{astro-ph/0703495}.

\bibitem[{\citenamefont{Sopuerta and Yunes}(2009)}]{Sopuerta:2009iy}
\bibinfo{author}{\bibfnamefont{C.~F.} \bibnamefont{Sopuerta}} \bibnamefont{and}
  \bibinfo{author}{\bibfnamefont{N.}~\bibnamefont{Yunes}},
  \bibinfo{journal}{Phys.Rev.} \textbf{\bibinfo{volume}{D80}},
  \bibinfo{pages}{064006} (\bibinfo{year}{2009}), \eprint{0904.4501}.

\bibitem[{\citenamefont{Canizares et~al.}(2012)\citenamefont{Canizares, Gair,
  and Sopuerta}}]{prisgair}
\bibinfo{author}{\bibfnamefont{P.}~\bibnamefont{Canizares}},
  \bibinfo{author}{\bibfnamefont{J.~R.} \bibnamefont{Gair}}, \bibnamefont{and}
  \bibinfo{author}{\bibfnamefont{C.~F.} \bibnamefont{Sopuerta}}
  (\bibinfo{year}{2012}), \eprint{1205.1253}.

\bibitem[{\citenamefont{Jackiw and Pi}(2003)}]{jackiw}
\bibinfo{author}{\bibfnamefont{R.}~\bibnamefont{Jackiw}} \bibnamefont{and}
  \bibinfo{author}{\bibfnamefont{S.~Y.} \bibnamefont{Pi}},
  \bibinfo{journal}{Phys. Rev.} \textbf{\bibinfo{volume}{D68}},
  \bibinfo{pages}{104012} (\bibinfo{year}{2003}), \eprint{gr-qc/0308071}.

\bibitem[{\citenamefont{Alexander and Yunes}(2009)}]{CSreview}
\bibinfo{author}{\bibfnamefont{S.}~\bibnamefont{Alexander}} \bibnamefont{and}
  \bibinfo{author}{\bibfnamefont{N.}~\bibnamefont{Yunes}},
  \bibinfo{journal}{Phys. Rept.} \textbf{\bibinfo{volume}{480}},
  \bibinfo{pages}{1} (\bibinfo{year}{2009}), \eprint{0907.2562}.

\bibitem[{\citenamefont{Polchinski}(1998)}]{polchinski2}
\bibinfo{author}{\bibfnamefont{J.}~\bibnamefont{Polchinski}},
  \emph{\bibinfo{title}{String theory. Vol. 2: Superstring theory and beyond}}
  (\bibinfo{publisher}{Cambridge University Press},
  \bibinfo{address}{Cambridge, UK}, \bibinfo{year}{1998}).

\bibitem[{\citenamefont{Alexander and Gates}(2006)}]{alexandergates}
\bibinfo{author}{\bibfnamefont{S.~H.~S.} \bibnamefont{Alexander}}
  \bibnamefont{and} \bibinfo{author}{\bibfnamefont{S.~J.} \bibnamefont{Gates},
  \bibfnamefont{Jr.}}, \bibinfo{journal}{JCAP} \textbf{\bibinfo{volume}{0606}},
  \bibinfo{pages}{018} (\bibinfo{year}{2006}), \eprint{hep-th/0409014}.

\bibitem[{\citenamefont{Taveras and Yunes}(2008)}]{taveras}
\bibinfo{author}{\bibfnamefont{V.}~\bibnamefont{Taveras}} \bibnamefont{and}
  \bibinfo{author}{\bibfnamefont{N.}~\bibnamefont{Yunes}},
  \bibinfo{journal}{Phys. Rev.} \textbf{\bibinfo{volume}{D78}},
  \bibinfo{pages}{064070} (\bibinfo{year}{2008}), \eprint{0807.2652}.

\bibitem[{\citenamefont{Calcagni and Mercuri}(2009)}]{calcagni}
\bibinfo{author}{\bibfnamefont{G.}~\bibnamefont{Calcagni}} \bibnamefont{and}
  \bibinfo{author}{\bibfnamefont{S.}~\bibnamefont{Mercuri}},
  \bibinfo{journal}{Phys. Rev.} \textbf{\bibinfo{volume}{D79}},
  \bibinfo{pages}{084004} (\bibinfo{year}{2009}), \eprint{0902.0957}.

\bibitem[{\citenamefont{Weinberg}(2008)}]{weinberg-CS}
\bibinfo{author}{\bibfnamefont{S.}~\bibnamefont{Weinberg}},
  \bibinfo{journal}{Phys. Rev.} \textbf{\bibinfo{volume}{D77}},
  \bibinfo{pages}{123541} (\bibinfo{year}{2008}), \eprint{0804.4291}.

\bibitem[{\citenamefont{Deser et~al.}(1982)\citenamefont{Deser, Jackiw, and
  Templeton}}]{deser-TMG}
\bibinfo{author}{\bibfnamefont{S.}~\bibnamefont{Deser}},
  \bibinfo{author}{\bibfnamefont{R.}~\bibnamefont{Jackiw}}, \bibnamefont{and}
  \bibinfo{author}{\bibfnamefont{S.}~\bibnamefont{Templeton}},
  \bibinfo{journal}{Annals Phys.} \textbf{\bibinfo{volume}{140}},
  \bibinfo{pages}{372} (\bibinfo{year}{1982}).

\bibitem[{\citenamefont{Grumiller and Yunes}(2008)}]{grumiller}
\bibinfo{author}{\bibfnamefont{D.}~\bibnamefont{Grumiller}} \bibnamefont{and}
  \bibinfo{author}{\bibfnamefont{N.}~\bibnamefont{Yunes}},
  \bibinfo{journal}{Phys. Rev.} \textbf{\bibinfo{volume}{D77}},
  \bibinfo{pages}{044015} (\bibinfo{year}{2008}), \eprint{0711.1868}.

\bibitem[{\citenamefont{Yunes and Sopuerta}(2008)}]{Yunes:2007ss}
\bibinfo{author}{\bibfnamefont{N.}~\bibnamefont{Yunes}} \bibnamefont{and}
  \bibinfo{author}{\bibfnamefont{C.~F.} \bibnamefont{Sopuerta}},
  \bibinfo{journal}{Phys. Rev.} \textbf{\bibinfo{volume}{D77}},
  \bibinfo{pages}{064007} (\bibinfo{year}{2008}), \eprint{0712.1028}.

\bibitem[{\citenamefont{{Yagi} et~al.}(2012)\citenamefont{{Yagi}, {Stein},
  {Yunes}, and {Tanaka}}}]{quadratic}
\bibinfo{author}{\bibfnamefont{K.}~\bibnamefont{{Yagi}}},
  \bibinfo{author}{\bibfnamefont{L.~C.} \bibnamefont{{Stein}}},
  \bibinfo{author}{\bibfnamefont{N.}~\bibnamefont{{Yunes}}}, \bibnamefont{and}
  \bibinfo{author}{\bibfnamefont{T.}~\bibnamefont{{Tanaka}}},
  \bibinfo{journal}{Phys. Rev.} \textbf{\bibinfo{volume}{D85}},
  \bibinfo{eid}{064022} (\bibinfo{year}{2012}), \eprint{1110.5950}.

\bibitem[{\citenamefont{Ali-Haimoud and Chen}(2011)}]{alihaimoud-chen}
\bibinfo{author}{\bibfnamefont{Y.}~\bibnamefont{Ali-Haimoud}} \bibnamefont{and}
  \bibinfo{author}{\bibfnamefont{Y.}~\bibnamefont{Chen}},
  \bibinfo{journal}{Phys.Rev.} \textbf{\bibinfo{volume}{D84}},
  \bibinfo{pages}{124033} (\bibinfo{year}{2011}), \eprint{1110.5329}.

\bibitem[{\citenamefont{Everitt et~al.}(2011)}]{GPB}
\bibinfo{author}{\bibfnamefont{C.~W.~F.} \bibnamefont{Everitt}}
  \bibnamefont{et~al.}, \bibinfo{journal}{Phys. Rev. Lett.}
  \textbf{\bibinfo{volume}{106}}, \bibinfo{pages}{221101}
  (\bibinfo{year}{2011}), \eprint{1105.3456}.

\bibitem[{\citenamefont{Ciufolini and Pavlis}(2004)}]{LAGEOS}
\bibinfo{author}{\bibfnamefont{I.}~\bibnamefont{Ciufolini}} \bibnamefont{and}
  \bibinfo{author}{\bibfnamefont{E.~C.} \bibnamefont{Pavlis}},
  \bibinfo{journal}{Nature} \textbf{\bibinfo{volume}{431}},
  \bibinfo{pages}{958} (\bibinfo{year}{2004}).

\bibitem[{\citenamefont{Iorio et~al.}(2011)\citenamefont{Iorio, Lichtenegger,
  Ruggiero, and Corda}}]{iorio-LT}
\bibinfo{author}{\bibfnamefont{L.}~\bibnamefont{Iorio}},
  \bibinfo{author}{\bibfnamefont{H.~I.} \bibnamefont{Lichtenegger}},
  \bibinfo{author}{\bibfnamefont{M.~L.} \bibnamefont{Ruggiero}},
  \bibnamefont{and} \bibinfo{author}{\bibfnamefont{C.}~\bibnamefont{Corda}},
  \bibinfo{journal}{Astrophys.Space Sci.} \textbf{\bibinfo{volume}{331}},
  \bibinfo{pages}{351} (\bibinfo{year}{2011}), \eprint{1009.3225}.

\bibitem[{\citenamefont{Kapner et~al.}(2007)\citenamefont{Kapner, Cook,
  Adelberger, Gundlach, Heckel et~al.}}]{kapner}
\bibinfo{author}{\bibfnamefont{D.}~\bibnamefont{Kapner}},
  \bibinfo{author}{\bibfnamefont{T.}~\bibnamefont{Cook}},
  \bibinfo{author}{\bibfnamefont{E.}~\bibnamefont{Adelberger}},
  \bibinfo{author}{\bibfnamefont{J.}~\bibnamefont{Gundlach}},
  \bibinfo{author}{\bibfnamefont{B.~R.} \bibnamefont{Heckel}},
  \bibnamefont{et~al.}, \bibinfo{journal}{Phys.Rev.Lett.}
  \textbf{\bibinfo{volume}{98}}, \bibinfo{pages}{021101}
  (\bibinfo{year}{2007}), \eprint{hep-ph/0611184}.

\bibitem[{\citenamefont{Zerilli}(1970)}]{zerilli}
\bibinfo{author}{\bibfnamefont{F.~J.} \bibnamefont{Zerilli}},
  \bibinfo{journal}{Phys. Rev.} \textbf{\bibinfo{volume}{D2}},
  \bibinfo{pages}{2141} (\bibinfo{year}{1970}).

\bibitem[{\citenamefont{Sago et~al.}(2003)\citenamefont{Sago, Nakano, and
  Sasaki}}]{sago}
\bibinfo{author}{\bibfnamefont{N.}~\bibnamefont{Sago}},
  \bibinfo{author}{\bibfnamefont{H.}~\bibnamefont{Nakano}}, \bibnamefont{and}
  \bibinfo{author}{\bibfnamefont{M.}~\bibnamefont{Sasaki}},
  \bibinfo{journal}{Phys. Rev.} \textbf{\bibinfo{volume}{D67}},
  \bibinfo{pages}{104017} (\bibinfo{year}{2003}), \eprint{gr-qc/0208060}.

\bibitem[{\citenamefont{Sopuerta and Yunes}(2011)}]{sopuertayunes}
\bibinfo{author}{\bibfnamefont{C.~F.} \bibnamefont{Sopuerta}} \bibnamefont{and}
  \bibinfo{author}{\bibfnamefont{N.}~\bibnamefont{Yunes}},
  \bibinfo{journal}{Phys. Rev.} \textbf{\bibinfo{volume}{D84}},
  \bibinfo{pages}{124060} (\bibinfo{year}{2011}), \eprint{1109.0572}.

\bibitem[{\citenamefont{{Benenti} and {Francaviglia}}(1979)}]{benenti}
\bibinfo{author}{\bibfnamefont{S.}~\bibnamefont{{Benenti}}} \bibnamefont{and}
  \bibinfo{author}{\bibfnamefont{M.}~\bibnamefont{{Francaviglia}}},
  \bibinfo{journal}{General Relativity and Gravitation}
  \textbf{\bibinfo{volume}{10}}, \bibinfo{pages}{79} (\bibinfo{year}{1979}).

\bibitem[{\citenamefont{Yasui and Houri}(2011)}]{yasui}
\bibinfo{author}{\bibfnamefont{Y.}~\bibnamefont{Yasui}} \bibnamefont{and}
  \bibinfo{author}{\bibfnamefont{T.}~\bibnamefont{Houri}},
  \bibinfo{journal}{Prog.Theor.Phys.Suppl.} \textbf{\bibinfo{volume}{189}},
  \bibinfo{pages}{126} (\bibinfo{year}{2011}), \eprint{1104.0852}.

\bibitem[{\citenamefont{Misner et~al.}(1973)\citenamefont{Misner, Thorne, and
  Wheeler}}]{MTW}
\bibinfo{author}{\bibfnamefont{C.~W.} \bibnamefont{Misner}},
  \bibinfo{author}{\bibfnamefont{K.}~\bibnamefont{Thorne}}, \bibnamefont{and}
  \bibinfo{author}{\bibfnamefont{J.~A.} \bibnamefont{Wheeler}},
  \emph{\bibinfo{title}{Gravitation}} (\bibinfo{publisher}{W. H. Freeman \&
  Co.}, \bibinfo{address}{San Francisco}, \bibinfo{year}{1973}).

\bibitem[{\citenamefont{Poisson}(2004)}]{toolkit}
\bibinfo{author}{\bibfnamefont{E.}~\bibnamefont{Poisson}},
  \emph{\bibinfo{title}{A relativist's toolkit : the mathematics of black-hole
  mechanics}} (\bibinfo{publisher}{Cambridge Univ. Pr.},
  \bibinfo{address}{Cambridge, UK}, \bibinfo{year}{2004}).

\bibitem[{\citenamefont{Thorne}(1980)}]{thorne-MM}
\bibinfo{author}{\bibfnamefont{K.~S.} \bibnamefont{Thorne}},
  \bibinfo{journal}{Rev. Mod. Phys.} \textbf{\bibinfo{volume}{52}},
  \bibinfo{pages}{299} (\bibinfo{year}{1980}).

\bibitem[{\citenamefont{Geroch}(1970)}]{geroch}
\bibinfo{author}{\bibfnamefont{R.~P.} \bibnamefont{Geroch}},
  \bibinfo{journal}{J. Math. Phys.} \textbf{\bibinfo{volume}{11}},
  \bibinfo{pages}{2580} (\bibinfo{year}{1970}).

\bibitem[{\citenamefont{Hansen}(1974)}]{hansen}
\bibinfo{author}{\bibfnamefont{R.~O.} \bibnamefont{Hansen}},
  \bibinfo{journal}{J. Math. Phys.} \textbf{\bibinfo{volume}{15}},
  \bibinfo{pages}{46} (\bibinfo{year}{1974}).

\bibitem[{\citenamefont{Stephani et~al.}(2003)\citenamefont{Stephani, Kramer,
  MacCallum, Hoenselaers, and Herlt}}]{stephani}
\bibinfo{author}{\bibfnamefont{H.}~\bibnamefont{Stephani}},
  \bibinfo{author}{\bibfnamefont{D.}~\bibnamefont{Kramer}},
  \bibinfo{author}{\bibfnamefont{M.}~\bibnamefont{MacCallum}},
  \bibinfo{author}{\bibfnamefont{C.}~\bibnamefont{Hoenselaers}},
  \bibnamefont{and} \bibinfo{author}{\bibfnamefont{E.}~\bibnamefont{Herlt}},
  \emph{\bibinfo{title}{Exact solutions of Einstein's field equations}}
  (\bibinfo{publisher}{Cambridge Univ. Pr.}, \bibinfo{address}{Cambridge, UK},
  \bibinfo{year}{2003}).

\bibitem[{\citenamefont{Campanelli et~al.}(2009)\citenamefont{Campanelli,
  Lousto, and Zlochower}}]{campanelli}
\bibinfo{author}{\bibfnamefont{M.}~\bibnamefont{Campanelli}},
  \bibinfo{author}{\bibfnamefont{C.~O.} \bibnamefont{Lousto}},
  \bibnamefont{and}
  \bibinfo{author}{\bibfnamefont{Y.}~\bibnamefont{Zlochower}},
  \bibinfo{journal}{Phys. Rev.} \textbf{\bibinfo{volume}{D79}},
  \bibinfo{pages}{084012} (\bibinfo{year}{2009}), \eprint{0811.3006}.

\bibitem[{\citenamefont{{Yagi, K., and Yunes, N., and Tanaka, T.}}({to be
  published})}]{kent-CSGW}
\bibinfo{author}{\bibnamefont{{Yagi, K., and Yunes, N., and Tanaka, T.}}},
  \emph{\bibinfo{title}{{}}} (\bibinfo{year}{{to be published}}).

\bibitem[{\citenamefont{Bardeen et~al.}(1972)\citenamefont{Bardeen, Press, and
  Teukolsky}}]{bardeen}
\bibinfo{author}{\bibfnamefont{J.~M.} \bibnamefont{Bardeen}},
  \bibinfo{author}{\bibfnamefont{W.~H.} \bibnamefont{Press}}, \bibnamefont{and}
  \bibinfo{author}{\bibfnamefont{S.~A.} \bibnamefont{Teukolsky}},
  \bibinfo{journal}{Astrophys. J.} \textbf{\bibinfo{volume}{178}},
  \bibinfo{pages}{347} (\bibinfo{year}{1972}).

\bibitem[{\citenamefont{Carter}(1968)}]{carter}
\bibinfo{author}{\bibfnamefont{B.}~\bibnamefont{Carter}},
  \bibinfo{journal}{Phys. Rev.} \textbf{\bibinfo{volume}{174}},
  \bibinfo{pages}{1559} (\bibinfo{year}{1968}).

\bibitem[{\citenamefont{Walker and Penrose}(1970)}]{walkerpenrose1970}
\bibinfo{author}{\bibfnamefont{M.}~\bibnamefont{Walker}} \bibnamefont{and}
  \bibinfo{author}{\bibfnamefont{R.}~\bibnamefont{Penrose}},
  \bibinfo{journal}{Commun. Math. Phys.} \textbf{\bibinfo{volume}{18}},
  \bibinfo{pages}{265} (\bibinfo{year}{1970}).

\bibitem[{\citenamefont{Levi-Civita}(1904)}]{levi-civita}
\bibinfo{author}{\bibfnamefont{T.}~\bibnamefont{Levi-Civita}},
  \bibinfo{journal}{Math. Ann.} \textbf{\bibinfo{volume}{59}},
  \bibinfo{pages}{383} (\bibinfo{year}{1904}).

\bibitem[{\citenamefont{Goldstein et~al.}(2002)\citenamefont{Goldstein, Poole,
  and Safko}}]{goldstein}
\bibinfo{author}{\bibfnamefont{H.}~\bibnamefont{Goldstein}},
  \bibinfo{author}{\bibfnamefont{C.}~\bibnamefont{Poole}}, \bibnamefont{and}
  \bibinfo{author}{\bibfnamefont{J.}~\bibnamefont{Safko}},
  \emph{\bibinfo{title}{Classical mechanics}}
  (\bibinfo{publisher}{Addison-Wesley}, \bibinfo{address}{San Francisco},
  \bibinfo{year}{2002}).

\bibitem[{\citenamefont{Glampedakis and Babak}(2006)}]{glampedakis}
\bibinfo{author}{\bibfnamefont{K.}~\bibnamefont{Glampedakis}} \bibnamefont{and}
  \bibinfo{author}{\bibfnamefont{S.}~\bibnamefont{Babak}},
  \bibinfo{journal}{Class. Quant. Grav.} \textbf{\bibinfo{volume}{23}},
  \bibinfo{pages}{4167} (\bibinfo{year}{2006}), \eprint{gr-qc/0510057}.

\bibitem[{\citenamefont{Vigeland et~al.}(2011)\citenamefont{Vigeland, Yunes,
  and Stein}}]{vigelandnico}
\bibinfo{author}{\bibfnamefont{S.}~\bibnamefont{Vigeland}},
  \bibinfo{author}{\bibfnamefont{N.}~\bibnamefont{Yunes}}, \bibnamefont{and}
  \bibinfo{author}{\bibfnamefont{L.}~\bibnamefont{Stein}},
  \bibinfo{journal}{Phys. Rev.} \textbf{\bibinfo{volume}{D83}},
  \bibinfo{pages}{104027} (\bibinfo{year}{2011}), \eprint{1102.3706}.

\bibitem[{\citenamefont{Papapetrou}(1953)}]{papapetrou}
\bibinfo{author}{\bibfnamefont{A.}~\bibnamefont{Papapetrou}},
  \bibinfo{journal}{Ann. Physik.} \textbf{\bibinfo{volume}{12}},
  \bibinfo{pages}{309} (\bibinfo{year}{1953}).

\bibitem[{\citenamefont{Blanchet}(2006)}]{blanchet-review}
\bibinfo{author}{\bibfnamefont{L.}~\bibnamefont{Blanchet}},
  \bibinfo{journal}{Living Rev. Rel.} \textbf{\bibinfo{volume}{9}},
  \bibinfo{pages}{4} (\bibinfo{year}{2006}).

\bibitem[{\citenamefont{Gair et~al.}(2008)\citenamefont{Gair, Li, and
  Mandel}}]{gairbumpy}
\bibinfo{author}{\bibfnamefont{J.~R.} \bibnamefont{Gair}},
  \bibinfo{author}{\bibfnamefont{C.}~\bibnamefont{Li}}, \bibnamefont{and}
  \bibinfo{author}{\bibfnamefont{I.}~\bibnamefont{Mandel}},
  \bibinfo{journal}{Phys. Rev.} \textbf{\bibinfo{volume}{D77}},
  \bibinfo{pages}{024035} (\bibinfo{year}{2008}), \eprint{0708.0628}.

\bibitem[{\citenamefont{Bambi and
  Barausse}(2011{\natexlab{a}})}]{bambibarausse-CZV}
\bibinfo{author}{\bibfnamefont{C.}~\bibnamefont{Bambi}} \bibnamefont{and}
  \bibinfo{author}{\bibfnamefont{E.}~\bibnamefont{Barausse}},
  \bibinfo{journal}{Phys.Rev.} \textbf{\bibinfo{volume}{D84}},
  \bibinfo{pages}{084034} (\bibinfo{year}{2011}{\natexlab{a}}),
  \eprint{1108.4740}.

\bibitem[{\citenamefont{Vigeland and Hughes}(2010)}]{vigelandhughes}
\bibinfo{author}{\bibfnamefont{S.~J.} \bibnamefont{Vigeland}} \bibnamefont{and}
  \bibinfo{author}{\bibfnamefont{S.~A.} \bibnamefont{Hughes}},
  \bibinfo{journal}{Phys. Rev.} \textbf{\bibinfo{volume}{D81}},
  \bibinfo{pages}{024030} (\bibinfo{year}{2010}), \eprint{0911.1756}.

\bibitem[{\citenamefont{{Manko} and {Novikov}}(1992)}]{mn}
\bibinfo{author}{\bibfnamefont{V.~S.} \bibnamefont{{Manko}}} \bibnamefont{and}
  \bibinfo{author}{\bibfnamefont{I.~D.} \bibnamefont{{Novikov}}},
  \bibinfo{journal}{Classical and Quantum Gravity}
  \textbf{\bibinfo{volume}{9}}, \bibinfo{pages}{2477} (\bibinfo{year}{1992}).

\bibitem[{\citenamefont{Apostolatos et~al.}(2009)\citenamefont{Apostolatos,
  Lukes-Gerakopoulos, and Contopoulos}}]{apostolatos}
\bibinfo{author}{\bibfnamefont{T.~A.} \bibnamefont{Apostolatos}},
  \bibinfo{author}{\bibfnamefont{G.}~\bibnamefont{Lukes-Gerakopoulos}},
  \bibnamefont{and}
  \bibinfo{author}{\bibfnamefont{G.}~\bibnamefont{Contopoulos}},
  \bibinfo{journal}{Phys. Rev. Lett.} \textbf{\bibinfo{volume}{103}},
  \bibinfo{pages}{111101} (\bibinfo{year}{2009}), \eprint{0906.0093}.

\bibitem[{\citenamefont{Lukes-Gerakopoulos
  et~al.}(2010)\citenamefont{Lukes-Gerakopoulos, Apostolatos, and
  Contopoulos}}]{gerakopoulos}
\bibinfo{author}{\bibfnamefont{G.}~\bibnamefont{Lukes-Gerakopoulos}},
  \bibinfo{author}{\bibfnamefont{T.~A.} \bibnamefont{Apostolatos}},
  \bibnamefont{and}
  \bibinfo{author}{\bibfnamefont{G.}~\bibnamefont{Contopoulos}},
  \bibinfo{journal}{Phys. Rev.} \textbf{\bibinfo{volume}{D81}},
  \bibinfo{pages}{124005} (\bibinfo{year}{2010}), \eprint{1003.3120}.

\bibitem[{\citenamefont{Contopoulos et~al.}(2011)\citenamefont{Contopoulos,
  Lukes-Gerakopoulos, and Apostolatos}}]{contopoulos}
\bibinfo{author}{\bibfnamefont{G.}~\bibnamefont{Contopoulos}},
  \bibinfo{author}{\bibfnamefont{G.}~\bibnamefont{Lukes-Gerakopoulos}},
  \bibnamefont{and} \bibinfo{author}{\bibfnamefont{T.~A.}
  \bibnamefont{Apostolatos}}, \bibinfo{journal}{Int. J. Bifurc. Chaos}
  \textbf{\bibinfo{volume}{21}}, \bibinfo{pages}{2261} (\bibinfo{year}{2011}),
  \eprint{1108.5057}.

\bibitem[{\citenamefont{Yunes and Pretorius}(2009{\natexlab{b}})}]{PPE}
\bibinfo{author}{\bibfnamefont{N.}~\bibnamefont{Yunes}} \bibnamefont{and}
  \bibinfo{author}{\bibfnamefont{F.}~\bibnamefont{Pretorius}},
  \bibinfo{journal}{Phys.Rev.} \textbf{\bibinfo{volume}{D80}},
  \bibinfo{pages}{122003} (\bibinfo{year}{2009}{\natexlab{b}}),
  \eprint{0909.3328}.

\bibitem[{\citenamefont{Yunes et~al.}(2010)\citenamefont{Yunes, Psaltis, Ozel,
  and Loeb}}]{Yunes:2009ch}
\bibinfo{author}{\bibfnamefont{N.}~\bibnamefont{Yunes}},
  \bibinfo{author}{\bibfnamefont{D.}~\bibnamefont{Psaltis}},
  \bibinfo{author}{\bibfnamefont{F.}~\bibnamefont{Ozel}}, \bibnamefont{and}
  \bibinfo{author}{\bibfnamefont{A.}~\bibnamefont{Loeb}},
  \bibinfo{journal}{Phys.Rev.} \textbf{\bibinfo{volume}{D81}},
  \bibinfo{pages}{064020} (\bibinfo{year}{2010}), \eprint{0912.2736}.

\bibitem[{\citenamefont{Yagi}(2012)}]{kent-LMXB}
\bibinfo{author}{\bibfnamefont{K.}~\bibnamefont{Yagi}} (\bibinfo{year}{2012}),
  \eprint{1204.4524}.

\bibitem[{\citenamefont{Johannsen and Psaltis}(2010)}]{johannsen-BHshadow}
\bibinfo{author}{\bibfnamefont{T.}~\bibnamefont{Johannsen}} \bibnamefont{and}
  \bibinfo{author}{\bibfnamefont{D.}~\bibnamefont{Psaltis}},
  \bibinfo{journal}{Astrophys. J.} \textbf{\bibinfo{volume}{718}},
  \bibinfo{pages}{446} (\bibinfo{year}{2010}), \eprint{1005.1931}.

\bibitem[{\citenamefont{Amarilla et~al.}(2010)\citenamefont{Amarilla, Eiroa,
  and Giribet}}]{amarilla}
\bibinfo{author}{\bibfnamefont{L.}~\bibnamefont{Amarilla}},
  \bibinfo{author}{\bibfnamefont{E.~F.} \bibnamefont{Eiroa}}, \bibnamefont{and}
  \bibinfo{author}{\bibfnamefont{G.}~\bibnamefont{Giribet}},
  \bibinfo{journal}{Phys. Rev.} \textbf{\bibinfo{volume}{D81}},
  \bibinfo{pages}{124045} (\bibinfo{year}{2010}), \eprint{1005.0607}.

\bibitem[{\citenamefont{Harko et~al.}(2010)\citenamefont{Harko, Kovacs, and
  Lobo}}]{harko}
\bibinfo{author}{\bibfnamefont{T.}~\bibnamefont{Harko}},
  \bibinfo{author}{\bibfnamefont{Z.}~\bibnamefont{Kovacs}}, \bibnamefont{and}
  \bibinfo{author}{\bibfnamefont{F.~S.~N.} \bibnamefont{Lobo}},
  \bibinfo{journal}{Class. Quant. Grav.} \textbf{\bibinfo{volume}{27}},
  \bibinfo{pages}{105010} (\bibinfo{year}{2010}), \eprint{0909.1267}.

\bibitem[{\citenamefont{Bambi and
  Barausse}(2011{\natexlab{b}})}]{bambibarausse-continuum}
\bibinfo{author}{\bibfnamefont{C.}~\bibnamefont{Bambi}} \bibnamefont{and}
  \bibinfo{author}{\bibfnamefont{E.}~\bibnamefont{Barausse}},
  \bibinfo{journal}{Astrophys.J.} \textbf{\bibinfo{volume}{731}},
  \bibinfo{pages}{121} (\bibinfo{year}{2011}{\natexlab{b}}),
  \eprint{1012.2007}.

\bibitem[{\citenamefont{Johannsen and Psaltis}(2011)}]{johannsen-QPO}
\bibinfo{author}{\bibfnamefont{T.}~\bibnamefont{Johannsen}} \bibnamefont{and}
  \bibinfo{author}{\bibfnamefont{D.}~\bibnamefont{Psaltis}},
  \bibinfo{journal}{Astrophys. J.} \textbf{\bibinfo{volume}{726}},
  \bibinfo{pages}{11} (\bibinfo{year}{2011}), \eprint{1010.1000}.

\bibitem[{\citenamefont{Johannsen and Psaltis}(2012)}]{johannsen-iron}
\bibinfo{author}{\bibfnamefont{T.}~\bibnamefont{Johannsen}} \bibnamefont{and}
  \bibinfo{author}{\bibfnamefont{D.}~\bibnamefont{Psaltis}}
  (\bibinfo{year}{2012}), \eprint{1202.6069}.

\bibitem[{\citenamefont{Chen and Jing}(2010)}]{chen}
\bibinfo{author}{\bibfnamefont{S.}~\bibnamefont{Chen}} \bibnamefont{and}
  \bibinfo{author}{\bibfnamefont{J.}~\bibnamefont{Jing}},
  \bibinfo{journal}{Class. Quant Grav.} \textbf{\bibinfo{volume}{27}},
  \bibinfo{pages}{225006} (\bibinfo{year}{2010}), \eprint{1005.1325}.

\bibitem[{grt()}]{grtensor}
\emph{\bibinfo{title}{{GRTensorII}}}, \bibinfo{note}{this is a package which
  runs within Maple but distinct from packages distributed with Maple. It is
  distributed freely on the World-Wide-Web from the address: {\tt
  http://grtensor.org}}.

\end{thebibliography}
\end{document}